\documentclass[aps,pre,superscriptaddress,floatfix,twocolumn,showpacs]{revtex4-1}

\usepackage{amsmath}
\usepackage{amsfonts}
\usepackage{amssymb}
\usepackage{graphicx}
\usepackage{bm}
\usepackage{color}
\usepackage{hyperref}
\usepackage{cases}
\usepackage{ulem}
\usepackage{multirow}
\usepackage{braket}

\newcommand{\1}{\mbox{1}\hspace{-0.25em}\mbox{l}}
\usepackage{bm}

\def\stacksymbols #1#2#3#4{\def\theguybelow{#2}
    \def\verticalposition{\lower#3pt}
    \def\spacingwithinsymbol{\baselineskip0pt\lineskip#4pt}
    \mathrel{\mathpalette\intermediary#1}}
\def\intermediary#1#2{\verticalposition\vbox{\spacingwithinsymbol
      \everycr={}\tabskip0pt
      \halign{$\mathsurround0pt#1\hfil##\hfil$\crcr#2\crcr
               \theguybelow\crcr}}}

\begin{document}
\title{Statistical properties of eigenvalues of the non-Hermitian\\ Su-Schrieffer-Heeger model with random hopping terms}

\author{Ken Mochizuki}
\affiliation{Department of Applied Physics, Hokkaido University, Sapporo 060-8628, Japan}
\author{Naomichi Hatano}
\affiliation{Institute of Industrial Science, University of Tokyo, Kashiwa 277-8574, Japan}
\author{Joshua Feinberg}
\affiliation{Department of Mathematics and Haifa Research Center for Theoretical Physics and
Astrophysics, University of Haifa, Mt. Carmel, Haifa 31905, Israel\\
https://orcid.org/0000-0002-2869-0010}
\author{Hideaki Obuse}
\affiliation{Department of Applied Physics, Hokkaido University, Sapporo 060-8628, Japan}
\affiliation{Institute of Industrial Science, University of Tokyo, Kashiwa 277-8574, Japan}

%\date{November 13, 2017}

\begin{abstract}
We explore the eigenvalue statistics of a non-Hermitian version of the  Su-Schrieffer-Heeger (SSH) model, with imaginary on-site potentials and randomly distributed hopping terms. We find that owing to the structure of the Hamiltonian, eigenvalues can be purely real in a certain range of parameters, even in the absence of Parity and Time-reversal symmetry. As it turns out, in this case of purely real spectrum, the level statistics is that of the Gaussian orthogonal ensemble. This demonstrates a general feature which we clarify that a non-Hermitian Hamiltonian whose eigenvalues are purely real can be mapped to a Hermitian Hamiltonian which inherits the symmetries of the original Hamiltonian. When the spectrum contains imaginary eigenvalues, we show that the density of states (DOS) vanishes at the origin and diverges at the spectral edges on the imaginary axis. We show that the divergence of the DOS originates from the Dyson singularity in chiral-symmetric one-dimensional Hermitian systems and derive analytically the asymptotes of the DOS which is different from that in Hermitian systems.
\end{abstract}

%\pacs{03.65.Vf, 03.67.−a, 05.30.Rt, 05.40.Fb}

\maketitle
\section{introduction}
\label{sec:introduction}
Non-Hermitian Hamiltonians have been studied extensively during the last couple of decades. This renewed interest was triggered mainly by Bender and Boettcher's discovery that non-Hermitian Hamiltonians with Parity and Time-reversal symmetry ($\mathcal{PT}$ symmetry) may have purely real spectra \cite{bender1998real}. (For a recent comprehensive review on $\mathcal{PT}$ symmetry see \cite{bender2019pt}.) Their work inspired numerous studies of non-Hermitian systems, not only theoretically but also experimentally, that explore foundations of non-Hermitian generalizations of quantum mechanics \cite{bender1999pt,bender2002complex,mostafazadeh2002pseudoI,mostafazadeh2002pseudoII,
mostafazadeh2002pseudoIII,mostafazadeh2004pseudounitary,
bender2007making,brody2013biorthogonal}, spontaneous $\mathcal{PT}$ symmetry breaking \cite{guo2009observation,zheng2010pt,ruter2010observation,miri2012large,regensburger2012parity,
chtchelkatchev2012stimulation,schomerus2013topologically,feng2014single,hodaei2014parity,
peng2014parity,peng2014loss,poli2015selective,zeuner2015observation,mochizuki2016explicit,
kim2016floquet,xiao2017observation}, non-Hermitian topological phases \cite{hu2011absence,esaki2011edge,
leykam2017edge,xiao2017observation,alvarez2018non,
shen2018topological,rudner2009topological,
weimann2017topologically,kunst2018biorthogonal,qi2018defect,
lieu2018topological,dangel2018topological,yao2018edge,
ghatak2019new,kawabata2019symmetry,borgnia2020non,xiao2020non}, to name a few. $\mathcal{PT}$ symmetry is fragile in its response to introducing arbitrary spatial disorder, unless the latter is introduced in a parity symmetric manner. In contrast, the more general class of pseudo-Hermitian systems \cite{mostafazadeh2002pseudoI,mostafazadeh2002pseudoII,mostafazadeh2002pseudoIII}, whose definition does not necessarily include the parity operation, are amenable to introducing disorder. \\\indent
Non-Hermitian systems with randomness have been studied in the context of  Anderson localization \cite{hatano1996localization,hatano1997vortex,hatano1998non,goldsheid1998distribution,
shnerb1998winding,feinberg1999non,feinberg1999curves, kolesnikov2000localization,moiseyev2001non,amir2016non,
jiang2019interplay,hamazaki2019non,zhang2020non}, low-energy QCD \cite{markum1999non,verbaarschot2000random,halasz1997fermion,halasz1997random,akemann2006unquenched,
akemann2007matrix,nishigaki2012level,nishigaki2012universality} and more \cite{feinberg2011effective,kalish2012light,hatano2016chebyshev,mochizuki2017effects}. The spectral statistics of random Hermitian Hamiltonians usually exhibits universal behavior, depending only on symmetries of the system \cite{bohigas1984characterization}. An interesting question then naturally arises whether the spectral statistics of disordered non-Hermitian Hamiltonians also exhibits universal behavior \cite{ginibre1965statistical,fyodorov1997almost,
chalker1998eigenvector,bernard2002classification,
shukla2001non,garcia2002critical,ahmed2003ensemble,
feinberg2006non,akemann2009gap,akemann2009wigner,
joglekar2011level,bohigas2013non,graefe2015random,
hamazaki2019threefold,tzortzakakis2020non}.\\\indent
In the present paper, we explore the spectral statistics of the non-Hermitian disordered Su-Schrieffer-Heeger (SSH) model. This model without disorder is one of the most vigorously studied non-Hermitian models, having its topological properties studied theoretically \cite{rudner2009topological,schomerus2013topologically,kunst2018biorthogonal,qi2018defect,
lieu2018topological,dangel2018topological} and experimentally \cite{poli2015selective,zeuner2015observation,weimann2017topologically}. In contrast, its spectral statistics has not received attention thus far. The purpose of this paper is to fill in this gap. We show that the spectrum in this model may be purely real in a certain parameter region, despite breaking of $\mathcal{PT}$ symmetry by disorder. Furthermore, by invoking pseudo-Hermiticity of this model, we show that its level statistics follows that of the Gaussian orthogonal ensemble (GOE) \cite{mehta2004random} when all eigenvalues are real. To this end we construct explicitly the generic similarity transformation from the non-Hermitian Hamiltonian with entirely real spectrum to a Hermitian Hamiltonian, which inherits the symmetries of the original Hamiltonian. Moreover, we find that the density of states (DOS) becomes singular and diverges in the presence of imaginary eigenvalues owing to the Dyson singularity \cite{dyson1953the}.\\\indent
This paper is organized as follows. In Sec. \ref{sec:model}, we
introduce the non-Hermitian SSH model with imaginary on-site potentials. We present in Sec. \ref{sec:properties}
properties of the eigenvalues, which we determine from the symmetries and structures of
the non-Hermitian Hamiltonian. Section \ref{sec:level_statistics_real}
is devoted to discussing the level statistics in the case of purely real spectrum after establishing a general argument on the inheritance of symmetries. Based on the properties of the model established in Sec. \ref{sec:properties},  we study the behavior of the DOS in the presence of imaginary eigenvalues in Sec. \ref{sec:dos}. Section \ref{sec:summary} gives a summary of our results.

\section{model}
\label{sec:model}
\begin{figure}[tb]
\begin{center}
\includegraphics[width=8.5cm]{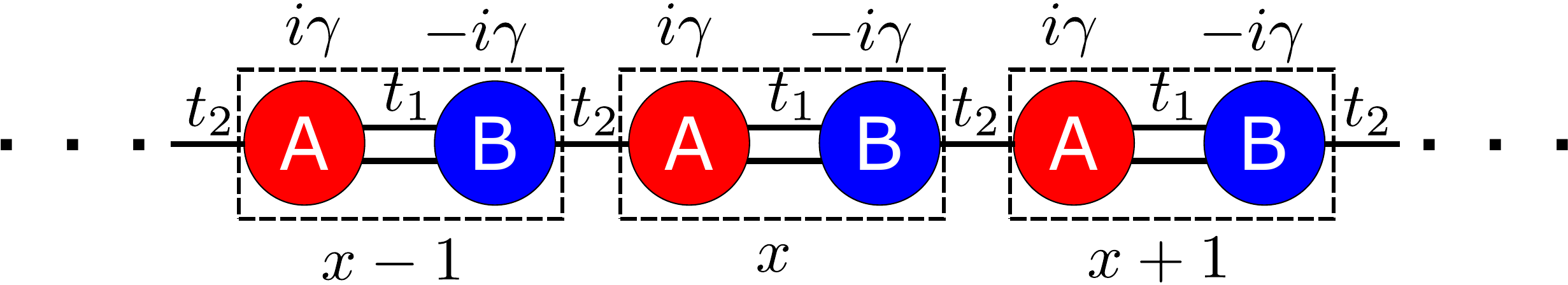}
\caption{The non-Hermitian SSH model. One unit-cell, enclosed by a dashed square, contains two sublattices $A$ (red) and $B$ (blue). While $\gamma$ is independent of $x$, $t_1(x)$ and $t_2(x)$ have random position-dependent values (which are suppressed in the figure for brevity). There are $N$  unit-cells in the chain, and we have imposed periodic boundary conditions in the numerical calculations.}
\label{fig:system}
\end{center}
\end{figure}

The non-Hermitian SSH model that we consider here, schematically depicted in Fig. \ref{fig:system}, is described by the Hamiltonian
\begin{align}
H=&H_0+H_\gamma,
\label{eq:total_Hamiltonian}\\
H_0=&\sum_x t_1(x)\ket{x,B}\bra{x,A}\nonumber\\
&\ \ \ \ +t_2(x)\ket{x+1,A}\bra{x,B}+\text{h.c.},
\label{eq:Hamiltonian_zero}\\ 
H_\gamma=&\sum_xi\gamma \ket{x,A}\bra{x,A}
-i\gamma \ket{x,B}\bra{x,B},
\label{eq:Hamiltonian_gamma}
\end{align}
where $A$ and $B$ are sublattice indices in the $x\,$th unit-cell. The system is comprised of $N$ such unit cells. Throughout this paper we assume periodic boundary conditions $x\sim x+N$, that is, the unit cells are arranged around a ring. The Hermitian term $H_0$, which consists of real intra- and inter-unit-cell hopping coefficients $\{t_1(x)\}$ and $\{t_2(x)\}$, is the Hamiltonian of the conventional Hermitian SSH model. The anti-Hermitian term $H_\gamma$, which makes $H$ non-Hermitian, describes on-site imaginary potentials $\pm i\gamma\ (\gamma\in\mathbb{R})$. In this basis, $H$ is a symmetric matrix with imaginary diagonal elements and real off-diagonal ones.  The present non-Hermitian SSH model can describe dynamics in single-mode waveguides or dielectric microwave resonators \cite{schomerus2013topologically,poli2015selective,zeuner2015observation,weimann2017topologically}. We can express a state $\ket{\phi}$ in the Hilbert space as
\begin{align}
\ket{\phi}=\sum_{x,\sigma}\phi_{\sigma}(x)\ket{x,\sigma},
\label{eq:wave_function}
\end{align}
where $\phi_\sigma(x)$ is the wave-function amplitude at the $\sigma\ (=A,B)$ sublattice in the $x\,$th unit-cell. 
\\\indent
The local hopping amplitudes $t_1(x)$ and $t_2(x)$ are identically and independently distributed, drawn from the box distributions
\begin{align}
t_{1/2}(x)\in[\bar{t}_{1/2}-w/2,\bar{t}_{1/2}+w/2],
\label{eq:random_hopping}
\end{align}
where the real parameters $\bar{t}_{1/2}$ and $w$ denote the mean values of
$t_{1/2}(x)$ and the width of the distribution, respectively. With no
loss of generality, we fix $\bar{t}_2=1$ in the following (and thereby set the scale of $t_1$ and $w$). In the case of no randomness $w=0$, $H$ is $\mathcal{PT}$ symmetric, namely
$(\mathcal{PT})H(\mathcal{PT})^{-1}=H$, with
$\mathcal{PT}=\sum_x\ket{-x}\bra{x}\sigma_1\mathcal{K}$, where
$\mathcal{K}$ is the complex conjugation operation and $\sigma_1$ is the appropriate standard Pauli matrix acting on the sublattice index. This $\mathcal{PT}$ symmetry renders all the eigenvalues real as long as $|t_1-t_2|>\gamma$, when $w=0$ \cite{kunst2018biorthogonal,qi2018defect,lieu2018topological,dangel2018topological}. In the case $w\neq0$, however, randomness of $t_1(x)$ and $t_2(x)$ breaks the $\mathcal{PT}$ symmetry of $H$. Nevertheless, in the next section we shall prove that the spectrum of the disordered $H$ may still be purely real, in some range of parameters.\\\indent
 We conclude this section with the remark that it is possible to generalize $\mathcal{PT}$ symmetry for our disordered system by taking $\mathcal{P}$ to be a certain unitary involution operator (not necessarily the parity operator) while maintaining $\mathcal{T}=\mathcal{K}$, such that this generalized $\mathcal{PT}$ symmetry ensures reality of the spectrum of $H$ in a certain range of parameters. However, there is no merit in doing so because the generalized $\mathcal{P}$ would depend on each random realization of $\{t_1(x)\}$ and $\{t_2(x)\}$; in other words, such a generalized $\mathcal{PT}$ symmetry would be lost under averaging over randomness.

\section{properties of eigenvalues}
\label{sec:properties}
In this section, we explain several interesting features of the complex eigenvalues of the {\it disordered} Hamiltonian $H$, which result from its symmetries and structure. Based on the classification made in Ref. \cite{kawabata2019symmetry}, our Hamiltonian has three symmetries: time-reversal symmetry
\begin{align}
H^{\rm T}=H,
\label{eq:time-reversal_symmetry}
\end{align}
particle-hole symmetry
\begin{align}
\tau_3H^\ast\tau_3=-H,
\label{eq:particle_hole_symmetry}
\end{align} 
and chiral symmetry
\begin{align}
\tau_3H\tau_3=-H^\dagger,
\label{eq:chiral_symmetry}
\end{align}
where $\tau_3=\1_x\otimes\sigma_3$ with $\sigma_3$ being a Pauli matrix and $\1_x=\sum_x\ket{x}\bra{x}$. In Eqs. (\ref{eq:time-reversal_symmetry})-(\ref{eq:chiral_symmetry}), $H^{\rm T}$, $H^\ast$, and $H^\dagger=(H^\ast)^{\rm T}$ respectively represent the transpose, complex conjugation, and Hermite conjugation of $H$. Particle-hole symmetry (or equivalently, chiral and time-reversal symmetry) implies that if $|E\rangle$ is an eigenstate with eigenvalue $E$, then $\tau_3 |E\rangle ^\ast$ is an eigenstate of $H$ with eigenvalue $-E^\ast$. Thus, eigenvalues $E$ with Re($E)\neq 0$ come in pairs $(E,-E^\ast)$, which are symmetric with respect to the imaginary axis. In fact, we shall show below that the eigenvalues of $H$ are either real or pure-imaginary. \\\indent
In order to investigate properties of eigenvalues in more detail, it is useful to write the eigenvalue equation for $H$ as
\begin{align}
H\left(\begin{array}{c}\ket{\alpha}\\\ket{\beta}\end{array}\right)
=E\left(\begin{array}{c}\ket{\alpha}\\\ket{\beta}\end{array}\right),
\label{eq:eigenvalue_equation_AB}
\end{align}
where $\ket{\alpha}$ and $\ket{\beta}$ represent wave functions on the two sublattices, as in $\ket{\alpha}=[\phi_A(1),\cdots,\phi_A(x),\cdots,\phi_A(N)]^\text{T}$ and $\ket{\beta}=[\phi_B(1),\cdots,\phi_B(x),\cdots,\phi_B(N)]^\text{T}$, respectively. In this basis, $H_0$ and $H_\gamma$ in Eqs. (\ref{eq:Hamiltonian_zero}) and (\ref{eq:Hamiltonian_gamma}) are given by
\begin{align}
H_0=\left(\begin{array}{cc}
0&\Omega\\
\Omega^\dagger&0
\end{array}\right),\ \ 
H_\gamma=i\gamma\tau_3,
\label{eq:Hamiltonian_AB}
\end{align}
where the $N\times N$ real, lower-triangular random matrix $\Omega$ is given by 
\begin{align}
\Omega_{x,x'}  = t_1(x)\delta_{x,x'}  + t_2(x-1) \delta_{x,x'+1}\,,
\label{eq:Omega}
\end{align}
and $\Omega^\dagger$ is of course its upper-triangular mirror image. For the reader's convenience, we display the unitary {\it chiral matrix}
\begin{align}
\tau_3=\left(\begin{array}{cc}
\1_x&0\\
0&-\1_x
\end{array}\right),
\label{eq:Gamma}
\end{align}
in this basis as well.  \\\indent
Let us digress briefly on the spectral properties of the Hermitian SSH Hamiltonian $H_0$ in Eq. (\ref{eq:Hamiltonian_AB}). The matrix $H_0$ anti-commutes with $\tau_3$:
\begin{align}
\{H_0,\tau_3\}=0.
\label{eq:anticommutation}
\end{align}
Thus, given an eigenstate $\ket{\tilde\psi_+}$
\begin{align}
H_0\ket{\tilde\psi_+} = H_0\left(\begin{array}{c}\ket{\tilde\alpha}\\\ket{\tilde\beta}\end{array}\right)
=E_0\left(\begin{array}{c}\ket{\tilde\alpha}\\\ket{\tilde\beta}\end{array}\right)
\label{eq:eigenvalue_equation_H0}
\end{align}
of $H_0$ with (real) eigenvalue $E_0$,
\begin{align}
\ket{\tilde\psi_-}  = \tau_3\ket{\tilde\psi_+}
\label{eq:chiralpartner}
\end{align}
is another eigenstate of $H_0$ corresponding to eigenvalue $-E_0$. The non-zero eigenvalues of $H_0$ come in pairs $\pm E_0$. 
Furthermore, it follows from Eq. (\ref{eq:eigenvalue_equation_H0}), for the components of $\ket{\tilde\psi_+}$, that 
\begin{align}
\Omega\Omega^\dagger\ket{\tilde{\alpha}}=E_0^2\ket{\tilde{\alpha}},\quad \Omega^\dagger\Omega\ket{\tilde{\beta}}=E_0^2\ket{\tilde{\beta}},
\label{eq:alpha-beta_tilde}
\end{align}
and
\begin{align}
\Omega^\dagger\ket{\tilde{\alpha}}=E_0\ket{\tilde{\beta}},\quad \Omega\ket{\tilde{\beta}}=E_0\ket{\tilde{\alpha}}.
\label{eq:SUSYmap}
\end{align}
Thus, the positive matrices $\Omega\Omega^\dagger$ and $\Omega^\dagger\Omega$ are isospectral. For $E_0\neq 0$, the respective eigenstates $\ket{\tilde\alpha}$ and $\ket{\tilde\beta}$ are related by Eq. (\ref{eq:SUSYmap}). For the components of $\ket{\tilde\psi_-}$, just flip the sign of $E_0$, or equivalently, of $\ket{\tilde\beta}$, in Eq. (\ref{eq:SUSYmap}). \\\indent Depending on the realization of disorder, $H_0$ may also have a doubly-degenerated zero eigenvalue $E_0=0$, with corresponding eigenstates 
\begin{align}
\ket{\tilde\psi_{0\pm}} = \left(\begin{array}{c}\ket{\tilde\alpha_0}\\\pm\ket{\tilde\beta_0}\end{array}\right),\ \  \ket{\tilde\psi_{0-}} = \tau_3\ket{\tilde\psi_{0+}}, 
\label{eq:zero-eigenvalue_H0}
\end{align}
where $\Omega \ket{\tilde\beta_0} =0$ and $\Omega^\dagger \ket{\tilde\alpha_0} =0$, with both $ \ket{\tilde\alpha_0} \neq 0$ and $\ket{\tilde\beta_0} \neq 0$. This is because $|E_0|$ is a singular value of $\Omega$ (and $\Omega^\dagger$) according to Eq. (\ref{eq:alpha-beta_tilde}). Thus, if $E_0 = 0$, $\Omega$ has a zero eigenvalue, with corresponding right- and left-eigenstates $\ket{\tilde\beta_0}$ and  $ \ket{\tilde\alpha_0}$, even if it is {\it not diagonalizable}. We can combine, of course,  the two null eigenstates in Eq. (\ref{eq:zero-eigenvalue_H0}) of $H_0$ into the two combinations $\ket{\tilde\psi_{0+}} \pm \ket{\tilde\psi_{0-}}$, which are simultaneous eigenstates of $\tau_3$. For periodic boundary conditions, this strict isospectrality of  $\Omega\Omega^\dagger$ and $\Omega^\dagger\Omega$, including zero modes (should they exist), persists also in the continuum limit (that is, $N\rightarrow\infty$, assuming the system makes a ring of some fixed length), corresponding to supersymmetric quantum mechanics \cite{dunne1998self}.  \\\indent
We shall now resume our discussion of the full non-Hermitian Hamiltonian $H$. From Eqs. (\ref{eq:total_Hamiltonian}), (\ref{eq:Hamiltonian_AB}), and (\ref{eq:anticommutation}), it follows that 
\begin{align}
H^2=\left(\begin{array}{cc}
\Omega\Omega^\dagger-\gamma^2\1_x&0\\
0&\Omega^\dagger\Omega-\gamma^2\1_x
\end{array}\right)
=H_0^2-\gamma^2\1,
\label{eq:square}
\end{align}
where $\1=\1_x\otimes\1_2$. Equation (\ref{eq:square}) means that the eigenvalues of $H$, $E\,(\in\mathbb{C})$, are given by the eigenvalues of $H_0$, $E_0\,(\in\mathbb{R})$, as in
\begin{align}
E=\pm \sqrt{E_0^2-\gamma^2}.
\label{eq:relation_E0-E}
\end{align}
Actually, up to overall normalization, we can write explicitly the eigenstates of $H$ in Eq. (\ref{eq:eigenvalue_equation_AB}), with eigenvalues $\pm \sqrt{E_0^2-\gamma^2}$, in terms of the corresponding eigenstates of $H_0$ with eigenvalues $\pm E_0$ in Eqs. (\ref{eq:eigenvalue_equation_H0})-(\ref{eq:chiralpartner}) as 
\begin{align}
\ket{\alpha}=\left(i\gamma\pm \sqrt{E_0^2-\gamma^2}\right)\ket{\tilde{\alpha}},\ 
\ket{\beta}=\Omega^\dagger\ket{\tilde{\alpha}}.
\label{eq:eigenstates}
\end{align}
Continuity at $\gamma=0$ (where $H$ coincides with $H_0$) means that the positive (negative) root in Eq. (\ref{eq:eigenstates}) gives the eigenstate of $H$ obtained from $\ket{\tilde\psi_+}\ (\ket{\tilde\psi_-})$. 

The doubly degenerate eigenvalue $E_0=0$ of $H_0$, should it exists, is split by the term $i\gamma\tau_3$ in $H$ into the pair of eigenvalues $E=\pm i\gamma$, which are the eigenvalues of $H$ with the largest and smallest imaginary parts. \\\indent 
\begin{figure}[bt]
\begin{center}
\includegraphics[width=9.5cm]{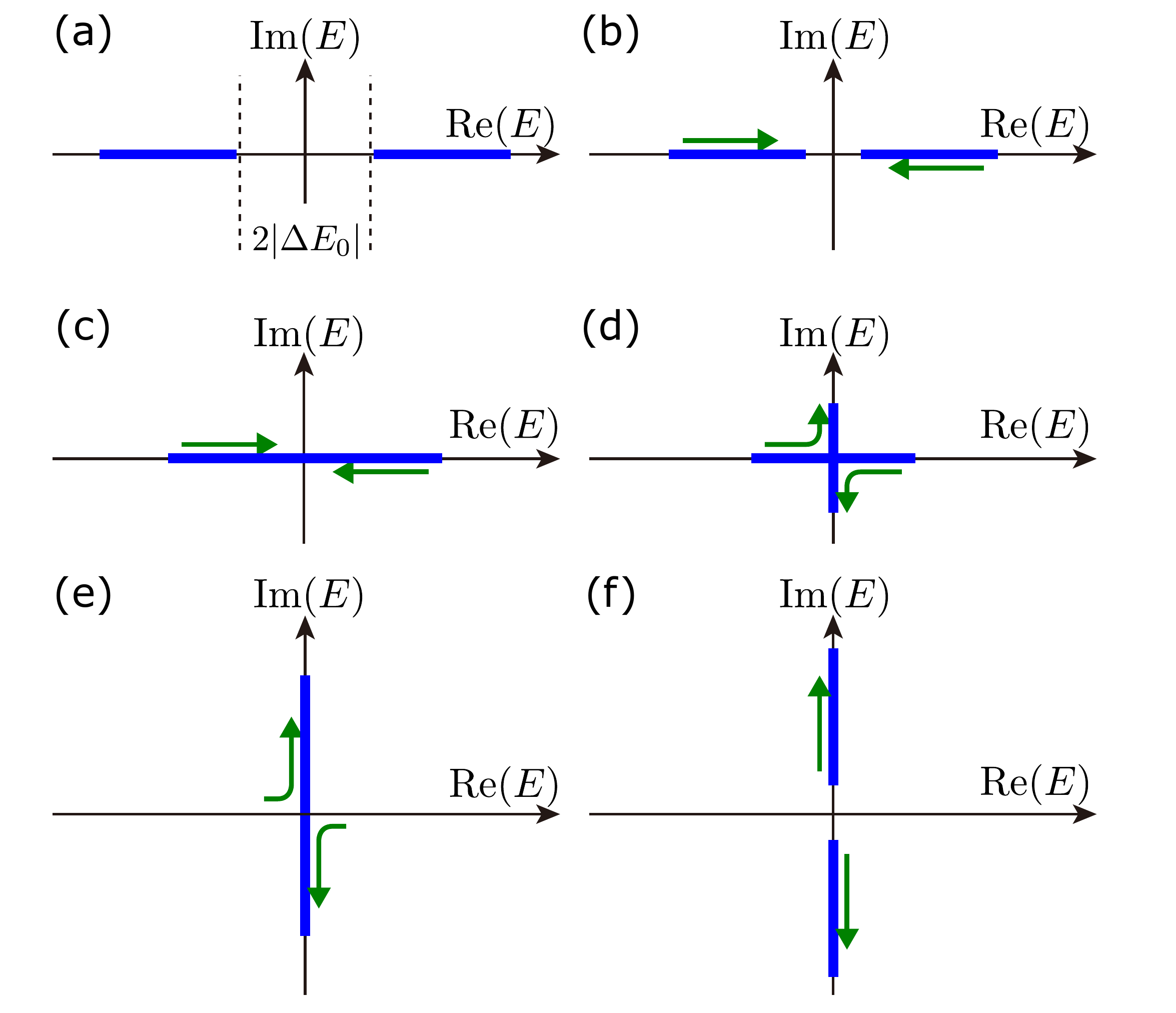}
\caption{Schematics that show the spectral change of the non-Hermitian SSH model in Eq. (\ref{eq:Hamiltonian_AB}) due to the increase of $\gamma$. Blue solid lines and green arrows represent eigenvalues and the direction to which eigenvalues shift with increasing $\gamma$, respectively. (a) The spectrum for the Hermitian case $\gamma=0$; series of eigenvalues on the real axis with a possible gap $\pm\Delta E_0$. (b) As we turn on $\gamma$, the gap around the origin is narrowed. (c) At the point $\Delta E_0=\gamma$, the gap closes. (d) The eigenvalues that reached the origin move onto the imaginary axis and away from the origin to up and down. (e) All eigenvalues are now on the imaginary axis. (f) A gap opens up on the imaginary axis.}
\label{fig:eigenvalue_shift}
\end{center}
\end{figure}
\begin{figure}[bt]
\begin{center}
\includegraphics[width=7cm]{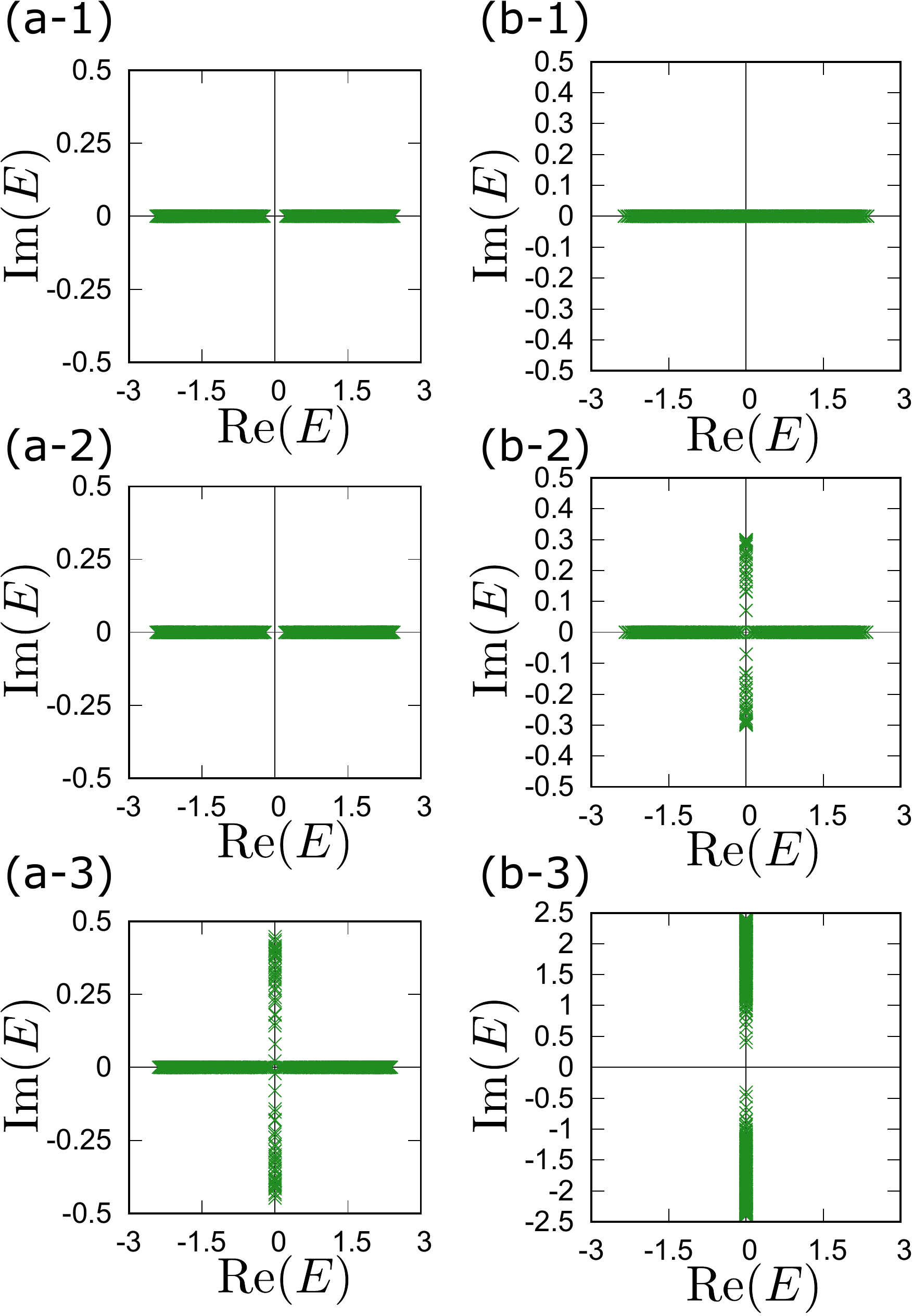}
\caption{Eigenvalues of $H$ when $N=360$ and periodic boundary conditions are imposed, with (a) $\bar{t}_1=1.3,\ \bar{t}_2=1.0,\ w=0.35$, and (b) $\bar{t}_1=1.0,\ \bar{t}_2=1.0,\ w=0.7$. In the left column, the values of $\gamma$ are (a-1) $\gamma=0.0$, (a-2) $\gamma=0.1$, and (a-3) $\gamma=0.5$. In the right column, $\gamma$ is varied as (b-1) $\gamma=0.0$, (b-2) $\gamma=0.3$, and (b-3) $\gamma=2.4$.}
\label{fig:eigenvalue}
\end{center}
\end{figure}
Since $H_0$ does not include $\gamma$, $E_0$ is independent of $\gamma$ and is determined only by $\{t_1(x)\}$ and $\{t_2(x)\}$. 
We can therefore understand the behavior of eigenvalues $E$ with increasing $\gamma$ in the following way. In order to give clear explanation, hereafter in this section, we use $E_n$ and $E_{0,n}$ for the $n$\,th eigenvalue of $H$ and $H_0$, respectively.
\begin{enumerate}
\item When $\gamma=0$ and hence $H=H_0$, the spectrum of the Hermitian SSH model is on the real axis and symmetric with respect to the origin, possibly with a gap around the origin as in Fig. \ref{fig:eigenvalue_shift} (a). As we turn on $\gamma$, eigenvalues $\pm |E_n|$ of $H$ on the real axis move toward the origin from right and left as in $E_n=\pm\sqrt{E_{0,n}^2-\gamma^2}$ and the gap around the origin $\pm\Delta E_0=\pm\min(|E_{0,n}|)$ at $\gamma=0$, if any, becomes narrower as in $\Delta E=\pm\sqrt{(\Delta E_0)^2-\gamma^2}$ [Fig. \ref{fig:eigenvalue_shift} (b)].
\item As $\gamma$ is increased, a pair of eigenvalues of the original values $\pm E_{0,n}$ meet at the origin when $E_{0,n}=\gamma$ [Fig. \ref{fig:eigenvalue_shift} (c)], and become pure imaginary as in $E_n=\pm i\sqrt{\gamma^2-E_{0,n}^2}$ [Fig. \ref{fig:eigenvalue_shift} (d)]. The point $E_{0,n}=\gamma$ is an exceptional point, where the two eigenstates become parallel to each other, and hence the matrix rank decreases by one.
\item The eigenvalues continue to move up and down on the imaginary axis as in $E_n=\pm i\sqrt{\gamma^2-E_{0,n}^2}$, which is shown in Fig. \ref{fig:eigenvalue_shift} (e). Increasing $\gamma$, all eigenvalues eventually move onto the imaginary axis, and then a gap $\pm i\sqrt{\gamma^2-\text{max}(E_{0,n})^2}$ opens up on the imaginary axis as in Fig. \ref{fig:eigenvalue_shift} (f). 
\end{enumerate}
Figure \ref{fig:eigenvalue} shows numerically obtained eigenvalues $E$
for the non-Hermitian SSH model with randomness. In Fig. \ref{fig:eigenvalue}, (a-1), (a-2), and (a-3)
correspond to (a), (b), and (d) in Fig. \ref{fig:eigenvalue_shift}. Even
when $\gamma\neq0$ and $H$ is non-Hermitian, all the eigenvalues of $H$
are real as long as the real line gap around $E=0$ exist, as shown in
Fig. \ref{fig:eigenvalue} (a-2). With increasing $\gamma$, the gap is
narrowed and pure imaginary eigenvalues appear after closing the gap
[Fig. \ref{fig:eigenvalue} (a-3)]. The right column, (b-1), (b-2), and
(b-3) in Fig. \ref{fig:eigenvalue} respectively correspond to (a), (b),
and (f) in Fig. \ref{fig:eigenvalue_shift} with $\Delta E_0=0$. In this
case, pure imaginary eigenvalues exist with any non-zero $\gamma$
[Fig. \ref{fig:eigenvalue} (b-2)]. As $\gamma$ is increased, all
eigenvalues become imaginary and the imaginary line gap is opened as
shown in Fig. \ref{fig:eigenvalue} (b-3). In both cases of $\Delta E_0\neq0$ and $\Delta E_0=0$, the imaginary part of $E_n$ cannot be larger than $|\gamma|$ and smaller than $-|\gamma|$, which results from Eq. (\ref{eq:relation_E0-E}).

\section{the level statistics when\\ all the eigenvalues are real}
\label{sec:level_statistics_real}
When the eigenvalues are entirely real as in Fig. \ref{fig:eigenvalue_shift} (b), we can show that the level statistics of the non-Hermitian SSH model obeys that of the GOE. To this end, we use a general fact that a non-Hermitian {\it diagonalizable} Hamiltonian $\mathcal{H}$ with entirely real eigenvalues can be transformed into a Hermitian Hamiltonian $\tilde{\mathcal{H}}$ by using a similarity transformation. Proving an inheritance of symmetries from $\mathcal{H}$ to $\tilde{\mathcal{H}}$, we discuss implications of these inherited symmetries for the level statistics of $H$. Then, we support our theoretical predictions by numerical simulations.
\subsection{inheritance of symmetries: general properties}
\label{sec:inheritance}
The real eigenspectrum of a diagonalizable non-Hermitian Hamiltonian $\mathcal{H}$ is determined by
\begin{align}
\mathcal{H}\ket{\psi_n}=E_n\ket{\psi_n},\ \mathcal{H}^\dagger\ket{\chi_n}=E_n\ket{\chi_n},\ E_n\in\mathbb{R},
\label{eq:eigenvalue_equation_LR}
\end{align}
where $\ket{\psi_n}$ and $\ket{\chi_n}$ are the right- and left-eigenstates corresponding to the real eigenvalue $E_n$. We henceforth assume a non-degenerate spectrum. The set of all these eigenstates comprises a bi-orthogonal basis, namely, they satisfy bi-orthonormality $\langle\chi_n|\psi_m\rangle=\delta_{nm}$ and completeness $\sum_n\ket{\psi_n}\bra{\chi_n}=\sum_n\ket{\chi_n}\bra{\psi_n}=\1$, where $\1$ is the identity operator. The spectral decomposition of $\mathcal{H}$ is 
\begin{align}
\mathcal{H}=\sum_n E_n \ket{\psi_n}\bra{\chi_n}.
\label{eq:spectral-decomposition-H}
\end{align}

In terms of these vectors, we can define a positive-definite Hermitian operator
\begin{align}
\eta=\sum_n\ket{\chi_n}\bra{\chi_n},
\label{eq:eta}
\end{align}
and its inverse
\begin{align}
\eta^{-1}=\sum_n\ket{\psi_n}\bra{\psi_n},
\label{eq:eta_inverse}
\end{align}
which transform $\mathcal{H}$ to $\mathcal{H}^\dagger$ \cite{mostafazadeh2002pseudoI,mostafazadeh2002pseudoII,mostafazadeh2002pseudoIII} as in
\begin{align}
\eta\mathcal{H}\eta^{-1}=\mathcal{H}^\dagger.
\label{eq:pseudo-Hermitian}
\end{align}
This property of $\mathcal{H}$, namely, that it is related to its Hermitian adjoint by a positive-definite similarity transformation, is sometimes referred to as pseudo-Hermiticity. (An alternative nomenclature is quasi-Hermiticity.)
We can thereby transform the non-Hermitian Hamiltonian $\mathcal{H}$ into a Hermitian Hamiltonian $\tilde{\mathcal{H}}$,
\begin{align}
\tilde{\mathcal{H}}=\eta^{1/2}\mathcal{H}\eta^{-1/2},
\label{eq:Hermitian_Hamiltonian}
\end{align}
where we used the fact that $\eta$ is Hermitian and positive-definite, from which it follows that the similarity transformation $\eta^{1/2}$ is Hermitian as well. We can confirm that $\tilde{\mathcal{H}}$ is Hermitian as in
\begin{align}
\tilde{\mathcal{H}}^\dagger=\eta^{-1/2}\mathcal{H}^\dagger\eta^{1/2}=\eta^{-1/2}\eta\mathcal{H}\eta^{-1}\eta^{1/2}=\tilde{\mathcal{H}},
\label{eq:Hermitian_proving}
\end{align}
which is ensured by Eq. (\ref{eq:pseudo-Hermitian}) and the Hermiticity of
$\eta^{\pm1/2}$. The similarity transformation in Eq. (\ref{eq:Hermitian_Hamiltonian}) implies that
$\mathcal{H}$ and $\tilde{\mathcal{H}}$ are isospectral.\\\indent
We can prove that the Hermitian Hamiltonian $\tilde{\mathcal{H}}$
inherits the symmetries of the non-Hermitian Hamiltonian $\mathcal{H}$
if all eigenvalues of $\mathcal{H}$ are real and not degenerate.
To this end, we first summarize the symmetries used in the classification
of non-Hermitian topological phases \cite{kawabata2019symmetry}. In the case of
Hermitian Hamiltonians,  time-reversal, particle-hole, and chiral symmetries are
defined as
\begin{align}
\mathcal{T} \tilde{\mathcal{H}}^* \mathcal{T}^{-1} &= \tilde{\mathcal{H}},
\label{eq:TRS_hermitian}\\
\mathcal{C} \tilde{\mathcal{H}}^* \mathcal{C}^{-1} &= -\tilde{\mathcal{H}},
\label{eq:PHS_hermitian}\\
\Gamma \tilde{\mathcal{H}} \Gamma^{-1} &= -\tilde{\mathcal{H}},
\label{eq:CS_hermitian}
\end{align}
respectively. The symmetry operators $\mathcal{T}, \mathcal{C}$, and $\Gamma$ are unitary operators which are constrained such that $\mathcal{T}\mathcal{T}^\ast$ and $\mathcal{C}\mathcal{C}^\ast$ are either $+\1$ or $-\1$, and $\Gamma^2=\1$. In the case of non-Hermitian Hamiltonians, the time-reversal and particle-hole symmetries ramify into two branches, namely AZ and AZ$^\dagger$
symmetries \cite{kawabata2019symmetry}, owing to the difference of transposition and complex conjugation for non-Hermitian Hamiltonians: $H^{\rm T}\neq H^\ast$. In the AZ symmetry class, they are defined as
\begin{align}
\mathcal{T} \mathcal{H}^* \mathcal{T}^{-1} &= \mathcal{H},
\label{eq:TRS}\\
\mathcal{C} \mathcal{H}^T \mathcal{C}^{-1} &= -\mathcal{H},
\label{eq:PHS}\\
\Gamma \mathcal{H} \Gamma^{-1} &= -\mathcal{H}^\dagger,
\label{eq:CS}
\end{align}
while in the AZ$^\dagger$ class they
are defined as
\begin{align}
\mathcal{T} \mathcal{H}^T \mathcal{T}^{-1} &= \mathcal{H},
\label{eq:TRS_dagger}\\
\mathcal{C} \mathcal{H}^* \mathcal{C}^{-1} &= -\mathcal{H},
\label{eq:PHS_dagger}\\
\Gamma \mathcal{H} \Gamma^{-1} &= -\mathcal{H}^\dagger.
\label{eq:CS_dagger}
\end{align}
In addition, the sublattice symmetry, which is equivalent to the chiral symmetry for Hermitian Hamiltonians, is now distinguished from the chiral symmetry because of the absence of Hermiticity $\mathcal{H}\neq\mathcal{H}^{\dagger}$,
\begin{align}
S \mathcal{H} S^{-1} &= -\mathcal{H},
\label{eq:SLS}
\end{align}
where $S$ is a unitary operator satisfying $S^2=\1$.\\\indent
 Next, we explain how the Hermitian Hamiltonian $\tilde{\mathcal{H}}$
inherits the symmetries of the non-Hermitian Hamiltonian $\mathcal{H}$. As a concrete example, we shall focus on time-reversal symmetry in the AZ$^{\dagger}$ class in Eq.\ (\ref{eq:TRS_dagger}) and the corresponding one in Hermitian case in
Eq.\ (\ref{eq:TRS_hermitian}). From Eqs. (\ref{eq:eigenvalue_equation_LR}) and (\ref{eq:TRS_dagger}), and from the assumption of non-degeneracy of the spectrum, we can see that $\ket{\psi_n}$ and $\ket{\chi_n}$ must satisfy
\begin{align}
\mathcal{T}\ket{\psi_n}^\ast=\kappa_n\ket{\chi_n}
\label{eq:psi_chi}
\end{align}
for real $E_n$, where $\kappa_n$ is a constant. We can choose the normalization constants of $\ket{\psi_n}$ and $\ket{\chi_n}$ such that $|\kappa_n|^2=1$. Equations (\ref{eq:eta}), (\ref{eq:eta_inverse}), and (\ref{eq:psi_chi}) then imply 
\begin{align}
\mathcal{T}(\eta^{-1})^\ast\mathcal{T}^{-1}=\eta,\ \
\mathcal{T}(\eta^\ast)^{-1/2} \mathcal{T}^{-1}=\eta^{1/2},
\label{eq:eta_relation}
\end{align}
because $\eta$ is Hermitian and positive definite. By using
Eqs. (\ref{eq:Hermitian_Hamiltonian}), (\ref{eq:Hermitian_proving}), (\ref{eq:TRS_dagger}), and (\ref{eq:eta_relation}), we can deduce that
\begin{align}
 \mathcal{T}\tilde{\mathcal{H}}^\ast\mathcal{T}^{-1}&=\mathcal{T}(\tilde{\mathcal{H}}^\dagger)^\ast\mathcal{T}^{-1}\nonumber\\
 &=\mathcal{T}(\eta^{-1/2})^\ast\mathcal{T}^{-1}\mathcal{T}\mathcal{H}^T\mathcal{T}^{-1}\mathcal{T}(\eta^{1/2})^\ast\mathcal{T}^{-1}\nonumber\\
 &=\eta^{1/2}\mathcal{H}\eta^{-1/2}=\tilde{\mathcal{H}},
\label{eq:real}
\end{align}
where we have also made use of  $\mathcal{T}\mathcal{T}^\ast=\pm\1$. Thus, we have proved that $\tilde{\mathcal{H}}$ satisfies
the relation for the time-reversal symmetry in Eq.\ (\ref{eq:TRS_hermitian}) with the same symmetry operator $\mathcal{T}$ of Eq.\ (\ref{eq:TRS_dagger}).
By following similar procedures, we can also prove inheritance of all symmetries in Eqs.\ (\ref{eq:TRS})-(\ref{eq:SLS}) from $\mathcal{H}$ to $\tilde{\mathcal{H}}$.

\begin{figure}[tbp]
\begin{center}
\includegraphics[width=8cm]{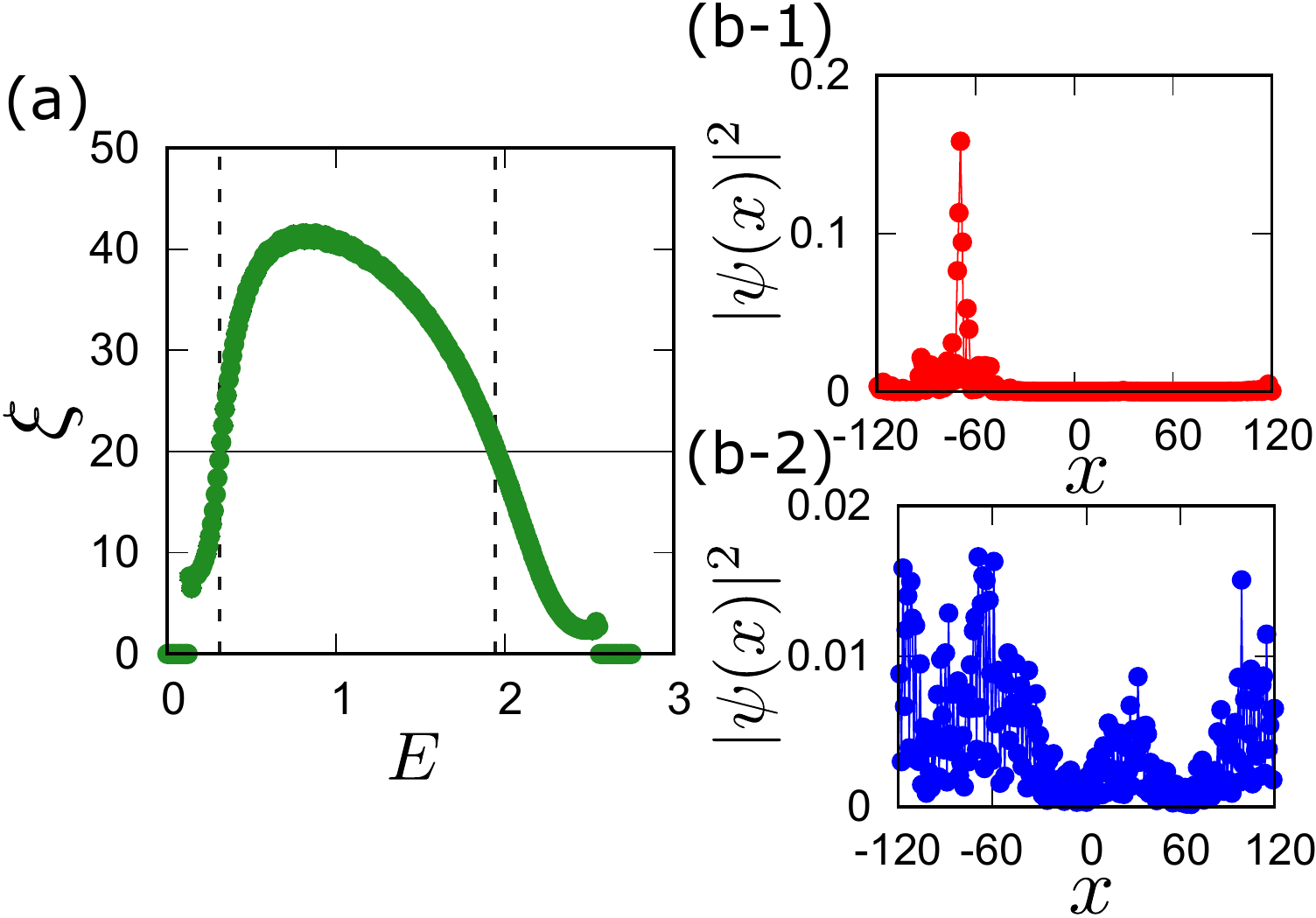}
\caption{(a) The localization length $\xi$ for the Hamiltonian $H$ with the same parameters as in Fig. \ref{fig:eigenvalue} (a-2): $\gamma=0.1,\ \bar{t}_1=1.3,\ \bar{t}_2=1.0$, and $w=0.35$. The system size is $N=240$ and the number of ensembles is $50\,000$. The horizontal solid line represents $\xi_c=20=N/12$. We take the data in the range of $E$ between the two vertical broken lines, namely, $0.31\leq E \leq1.94$. (b) Examples of eigenstates $|\psi(x)|^2=|\psi_A(x)|^2+|\psi_B(x)|^2$. (b-1) An eigenstate with $E\simeq2.10$, which is regarded as a localized state. (b-2) An eigenstate with $E\simeq1.29$, which is regarded as an extended state.}
\label{fig:localized_extended}
\end{center}
\end{figure}
\subsection{numerical confirmation}
\label{sec:confirmation}
Now, we focus on the non-Hermitian SSH model $H$ in Eq.\
(\ref{eq:total_Hamiltonian}). From Eqs. (\ref{eq:time-reversal_symmetry})-(\ref{eq:chiral_symmetry}) and (\ref{eq:TRS_dagger})-(\ref{eq:CS_dagger}), we infer that the non-Hermitian Hamiltonian $H$ belongs to the BDI$^\dagger$ class, with $\mathcal{T}=\1=\1_x\otimes\1_2,\ \mathcal{C}=\tau_3=\1_x\otimes\sigma_3$, and $\Gamma=\tau_3=\1_x\otimes\sigma_3$.
As long as all eigenvalues are real, $H$ can be transformed into the Hermitian Hamiltonian $\tilde{H}$ 
by Eq.\ (\ref{eq:Hermitian_Hamiltonian}).
As explained above, $\tilde{H}$ inherits and retains the symmetries of $H$, namely, time-reversal,
particle-hole, and chiral symmetries. In particular, time-reversal symmetry of $H$ in Eq.\
(\ref{eq:time-reversal_symmetry}) implies $\tilde{H}^*=\tilde{H}$, which means that $\tilde{H}$ is a real symmetric matrix. It is known that the
level statistics of real symmetric random matrices obeys that of the GOE when
eigenstates are extended \cite{bohigas1984characterization}. Therefore, as long as all eigenvalues of
$H$ are real, the level statistics of $H$ obeys that of the GOE as well, when
we focus on extended eigenstates whose eigenvalues are not too close to the origin.\\\indent
\begin{figure}[tbp]
\begin{center}
\includegraphics[width=6cm]{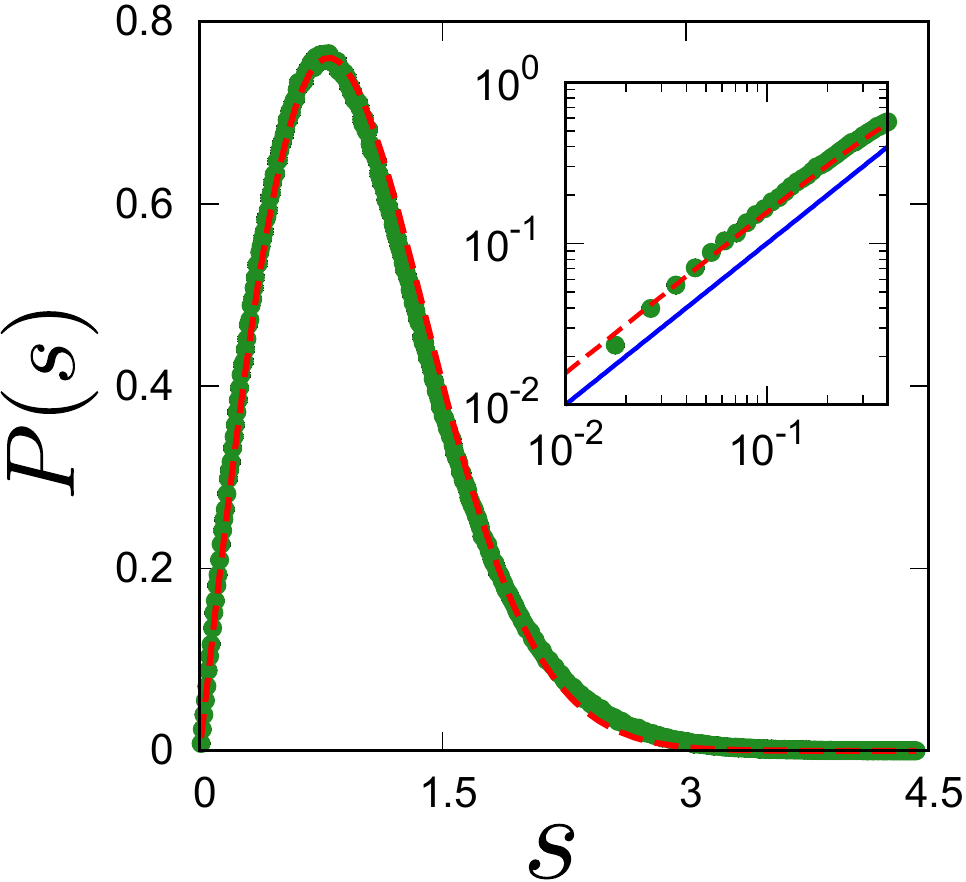}
\caption{The level-spacing distribution $P(s)$ are plotted as green dots with $\gamma=0.1,\ \bar{t}_1=1.3,\ \bar{t}_2=1.0$, and $w=0.35$,
 corresponding to the parameters in Fig. \ref{fig:eigenvalue} (a-2). The system size is $N=240$ and the number of ensembles is $50000$. The red broken line indicates the level-spacing distribution of the GOE in Eq.\ (\ref{eq:P(s)}). In the inset, $P(s)$ near $s=0$ is depicted in a logarithmic scale, where the blue solid line indicates $P(s) \propto s$.}
\label{fig:GOE}
\end{center}
\end{figure}
Here, we confirm this conclusion by numerical calculation of the level-spacing distribution. The present system is a one-dimensional random system and hence almost all eigenstates should be localized in the infinite system. However, in
finite systems, eigenstates whose localization lengths are comparable to the system size can be regarded as extended states. We therefore evaluate the eigenvalue dependence of the localization length $\xi$ in order to find the $E$ range of extended eigenstates. To this end, we assume exponential localization of the eigenstate as
\begin{align}
 \psi_\sigma(x) \propto \exp{\left(-\frac{|x|}{\xi}\right)},
\label{eq:xi}
\end{align}
where $\psi_\sigma(x)$ represents the wave-function amplitude at the sublattice $\sigma=A,B$ in the $x\,$th unit cell of the right eigenstate $\ket{\psi}$. Using Eq.\ (\ref{eq:xi}) and assuming $N\rightarrow \infty$, $\xi$ is calculated by
\begin{align}
\xi=\frac{I_1^2}{4I_2},
\label{eq:participation_ratio}
\end{align}
where the inverse participation ratio $I_m\ (m=1,2)$ is defined as
\begin{align}
I_m=\sum_{x,\sigma}|\psi_\sigma(x)|^{2m}.
\label{eq:I_m}
\end{align}
We note that $\xi$ defined in Eq.\ (\ref{eq:participation_ratio}) does not depend on the normalization constant.
\\\indent
Figure \ref{fig:localized_extended} shows the results of numerical calculations with
the same parameters as in Fig. \ref{fig:eigenvalue} (a-2). In
Fig. \ref{fig:localized_extended} (a), the localization length $\xi$ is plotted
as a function of $E$. In order to take eigenstates which can
be regarded as extended states, we focus on the range of $E$ in which
$\xi\ge\xi_c=20=N/12$ is satisfied, as shown in Fig. \ref{fig:localized_extended} (a). Two examples of eigenstates in an ensemble given in Fig. \ref{fig:localized_extended} (b) indeed show that one out of the range in (b-1) is regarded as a localized state and the other in the range in (b-2) is regarded as an extended state. We obtained the level-spacing distribution $P(s)$ from all
eigenvalues in this range of $E$ for 50\,000 samples. The normalized level spacing $s$ is defined
as $s=\delta E/\langle \delta E \rangle$, where $\delta E$ is the absolute difference of adjacent eigenvalues and $\langle \delta E \rangle$ is the mean value of $\delta E$ averaged over ensembles and the range of $E$. The obtained level-spacing distribution $P(s)$ 
agrees well with that of the GOE \cite{mehta2004random},
\begin{align}
 P(s)=\frac{\pi s}{2}\exp(-\pi s^2/4),
 \label{eq:P(s)}
\end{align}
as is observed in Fig. \ref{fig:GOE}. We thereby confirm that the level statistics of the non-Hermitian SSH model whose eigenvalues are entirely real obeys that of the GOE.\\\indent
 We note that, the strength of randomness $w$ should be set to an intermediate value. On one hand, if the value of $w$ were too large, all of the eigenstates would be localized even in finite systems and the level statistics would be the Poisson distribution. On the other hand, if the value of $w$ were too small, the level statistics would be similar to that of the clean system without randomness.

\begin{figure}[b]
\begin{center}
\includegraphics[width=8cm]{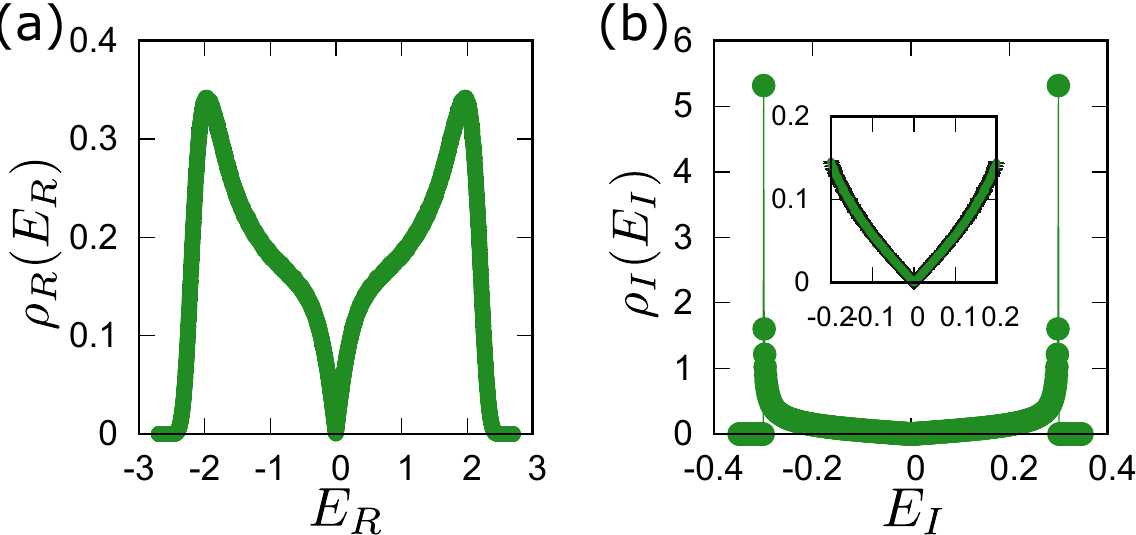}
\caption{Green dots represent the DOS for the Hamiltonian $H$ with $\gamma=0.3,\ \bar{t}_1=1.0,\ \bar{t}_2=1.0,$ and $w=0.7$, corresponding to the parameters in Fig. \ref{fig:eigenvalue} (b-2). Here, (a) $\rho_R(E_R)$ denotes the DOS of the eigenvalues on the real axis $E_R$, while (b) $\rho_I(E_I)$ denotes the one on the imaginary axis $E_I$ where the inset is the DOS near the origin. The system size is $N=300$ and the number of ensembles is $200000$. Note that, the bin size for the plot of the DOS in Fig. \ref{fig:DOS_RI_total} is $10^{-3}$ for both $E_R$ and $E_I$, different from that in Figs. \ref{fig:DOS_I_divergence} and \ref{fig:DOS_RI_vanish}.}
\label{fig:DOS_RI_total}
\end{center}
\end{figure}

\section{the DOS when pure imaginary eigenvalues exist}
\label{sec:dos}
 In this section, we study spectral properties in the case that $H_0$ is gapless ($\Delta E_0=0$) and eigenvalues of $H$ partially become pure imaginary at a finite value of $\gamma$, corresponding to Fig. \ref{fig:eigenvalue_shift} (d) and Fig. \ref{fig:eigenvalue} (b-2), thereby the argument in Sec. \ref{sec:level_statistics_real} cannot be applied. We focus on the DOS choosing the parameters so that the spectrum of $H_0$ can become gapless at $E_0=0$ ($\Delta E_0=0$). Figure \ref{fig:DOS_RI_total} is the DOS obtained numerically: Figure \ref{fig:DOS_RI_total} (a) and (b) respectively show the DOS of the eigenvalues on the real and imaginary axes, $\rho_R(E_R)$ and $\rho_I(E_I)$, by taking into account the fact that eigenvalues are either real $E_R\, (\in \mathbb{R})$ or pure imaginary $i E_I\, (\in i \mathbb{R})$, as explained in
Sec.\ \ref{sec:properties}. The DOS is normalized in the whole range of $E$ as in
\begin{align}
 \int_{-\infty}^{\infty} \rho_R(E_R) dE_R +  \int_{-\gamma}^{\gamma} \rho_I(E_I) dE_I =1. 
 \label{eq:normalization}
\end{align}
We find that the DOS exhibits two distinct features.
First,  $\rho_I(E_I)$ diverges at $E_I=\pm\gamma$ in
Fig. \ref{fig:DOS_RI_total} (b). 
Second, both $\rho_R(E_R)$ and $\rho_I(E_I)$ vanish at the origin as shown in Fig. \ref{fig:DOS_RI_total} (a) and (b), respectively.\\\indent
\begin{figure}[tb]
\begin{center}
\includegraphics[width=6cm]{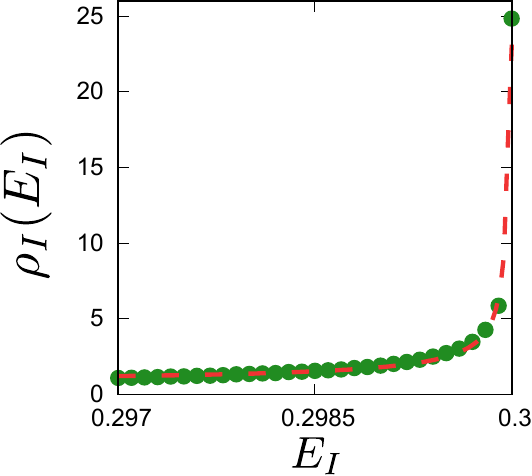}
\caption{The DOS $\rho_I(E_I)$ with the same parameters as Fig. \ref{fig:DOS_RI_total} are plotted as green dots near $E_I=\gamma=0.3$ where the bin size is $10^{-4}$. The red broken line represents $\rho_I(E_I)=\mu|E_I|/(\lambda^2-E_I^2)|\ln(\nu\sqrt{\lambda^2-E_I^2})|^3$, where fitting parameters are $\mu=0.0630\pm0.0013,\ \nu=3.04\pm0.07$, and $\lambda=0.29999\pm0.00023$. While the value of $\lambda$ should be $\gamma=0.3$, it is slightly shifted from $0.3$ due to finite size effects.}
\label{fig:DOS_I_divergence}
\end{center}
\end{figure}
The discussion in Sec. \ref{sec:properties} explains these features. Using the DOS of $H_0$, $\tilde{\rho}(E_0)$, we write the DOS of $H$ as
\begin{align}
\rho_{R/I}(E_{R/I})=\tilde{\rho}(E_0)\left|\frac{dE_0}{dE_{R/I}}\right|.
\label{eq:DOS}
\end{align}
According to Eqs. (\ref{eq:relation_E0-E}) and (\ref{eq:DOS}), we can derive relations
\begin{align}
\rho_R(E_R)&=\tilde{\rho}\left(\sqrt{E_R^2+\gamma^2}\right)\frac{|E_R|}{\sqrt{E_R^2+\gamma^2}},
\label{eq:DOS_R}\\
\rho_I(E_I)&=\tilde{\rho}\left(\sqrt{\gamma^2-E_I^2}\right)\frac{|E_I|}{\sqrt{\gamma^2-E_I^2}}.
\label{eq:DOS_I}
\end{align}
We elaborate the above features by using Eqs.\ (\ref{eq:DOS_R}) and (\ref{eq:DOS_I}). First, the divergence of $\rho_I(E_I)$ at $E_I=\pm \gamma$ originates
from the Dyson singularity of the DOS of the Hermitian Hamiltonian $H_0$ in Eq. (\ref{eq:Hamiltonian_AB}), which has chiral symmetry
\begin{align}
\tau_3 H_0 \tau_3=-H_0.
\label{eq:chiral-symmetry_Hermitian}
\end{align}
In chirally symmetric one-dimensional Hermitian systems, when we adjust the system parameters so that the spectrum of $H_0$ may be gapless at $E_0=0$, the DOS diverges at the gapless point. This is known as the Dyson singularity, whose functional form is
\begin{align}
\tilde{\rho}(E_0)=\frac{\mu}{|E_0[\ln(\nu|E_0|)]^3|},
\label{eq:divergence_E0}
\end{align}
around $E_0=0$ \cite{dyson1953the,theodorou1976extended,eggarter1978singular,titov2001fokker}, where the parameters $\mu$ and $\nu$ depend on details of the system. As was shown in Sec. \ref{sec:properties}, the point $E_0=0$ in the spectrum of $H_0$ is shifted to the points $E=\pm i\gamma$ in the spectrum of $H$. Hence the Dyson singularity, the divergence of $\tilde{\rho}(E_0=0)$, produces the divergence of $\rho_I(\pm\gamma)$ of the form
\begin{align}
\rho_I(E_I)=\frac{\mu|E_I|}{(\gamma^2-E_I^2)|\ln(\nu\sqrt{\gamma^2-E_I^2})|^3},
\label{eq:divergence_EI}
\end{align}
on the imaginary axis, which we find using Eqs. (\ref{eq:DOS_I}) and (\ref{eq:divergence_E0}). Figure \ref{fig:DOS_I_divergence} shows $\rho_I(E_I)$ which we numerically obtained. The numerical result agrees well with Eq. (\ref{eq:divergence_EI}). Next, for $E\simeq0$, we have
\begin{align}
\rho_{R/I}(E_{R/I})\simeq\tilde{\rho}(\gamma)\left|\frac{E_{R/I}}{\gamma}\right|.
\label{eq:near_origin}
\end{align}
Since there is no other divergences in $\tilde{\rho}(E_0)$ except at the origin,
both $\rho_R(E_R)$ and $\rho_I(E_I)$ vanish linearly as $|E_{R/I}|$, which is numerically demonstrated in Fig. \ref{fig:DOS_RI_vanish}.

\begin{figure}[bt]
\begin{center}
\includegraphics[width=8cm]{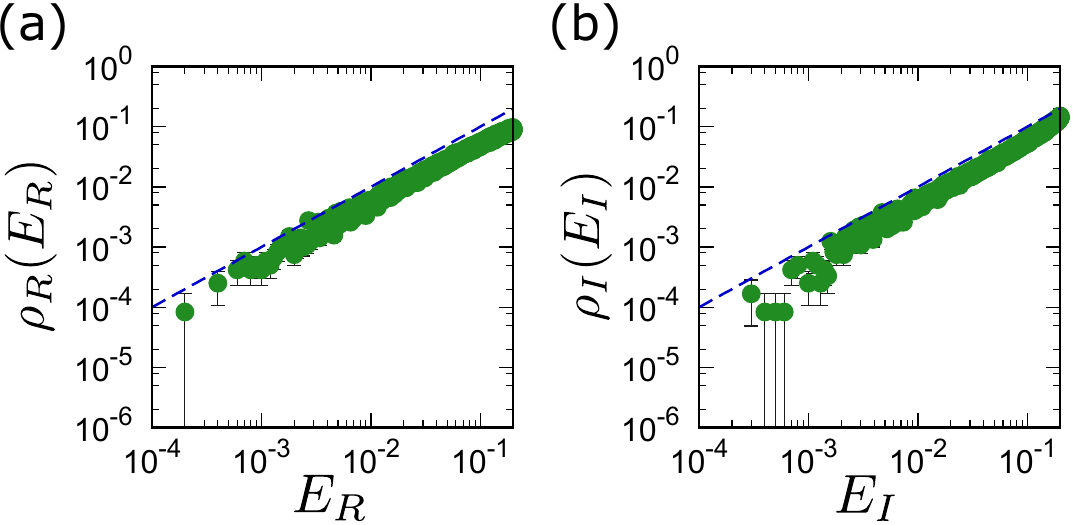}
\caption{The DOS around  $E=0$ with the same parameters and bin sizes as Fig. \ref{fig:DOS_I_divergence}. (a) $\rho_R(E_R)$ on the real axis $E_R$ and (b) $\rho_I(E_I)$ on the imaginary axis $E_I$ are plotted as green dots. The blue broken lines indicate $\rho_{R/I}(E_{R/I})=|E|$.}
\label{fig:DOS_RI_vanish}
\end{center}
\end{figure}

\section{summary}
\label{sec:summary}
We have explored statistical properties of eigenvalues of a non-Hermitian SSH model with randomly distributed hopping terms. This model may describe experimental settings in which single mode waveguides or dielectric microwave resonators with gain and loss are randomly arranged on a line \cite{schomerus2013topologically,poli2015selective,zeuner2015observation,weimann2017topologically}. We have proved that the eigenvalues of $H$ can be entirely real in the absence of $\mathcal{PT}$ symmetry owing to the structure of the Hamiltonian.\\\indent
Furthermore, we have shown that the level statistics of the effectively extended
eigenstates obeys that of the GOE when all eigenvalues are real. This is so because in this case the Hamiltonian $H$ is mapped to a Hermitian Hamiltonian
$\tilde{H}$ by positive-definite similarity transformation. Thus, $\tilde{H}$ inherits
all symmetries of $H$ when the eigenvalues of the non-Hermitian
Hamiltonian $H$ are entirely real. The latter statement which we have shown in Sec. \ref{sec:inheritance} is generic, and our model reaffirms its veracity as a particular example. \\\indent
We have also studied the DOS when pure imaginary eigenvalues
exist. There are two distinct features:
First, the DOS increases along the imaginary axis and diverges at
$E=\pm i \gamma$,
and second, the DOS decreases linearly toward the origin and vanishes at
$E=0$.
We have explained both features using a relation between the Hermitian Hamiltonian $H_0$ and the non-Hermitian Hamiltonian $H$. In particular, we have demonstrated that the Dyson singularity of the DOS of the Hermitian system due to its  chiral symmetry is the reason for the divergence of the DOS of the non-Hermtian $H$ at $E=\pm i\gamma$. This is the first study discussing the Dyson singularity in non-Hermitian systems. It should be interesting to explore other singularities in non-Hermitian random systems.

\section{acknowledgement}
\label{sec:acknowledgement}
This collaboration was initiated when the authors met at the International Centre for Theoretical Sciences (ICTS) during a visit for participating in the program `Non-Hermitian Physics - PHHQP XVIII' (Code: ICTS/nhp2018/06). We thank Yasuhiro Asano, Roman Riser, and Kousuke Yakubo for helpful discussions. The work of KM, NH and HO was supported by KAKENHI (Grants Nos. JP18J20727, JP18H01140, JP18K18733, JP19K03646, and JP19H00658). JF's work was supported by the Israel Science Foundation (ISF) under grant No. 2040/17.
%JF also thanks Roman Riser for a useful discussion. 

\bibliographystyle{apsrev4-1}
\bibliography{reference.bib}

\begin{thebibliography}{89}%
\makeatletter
\providecommand \@ifxundefined [1]{%
 \@ifx{#1\undefined}
}%
\providecommand \@ifnum [1]{%
 \ifnum #1\expandafter \@firstoftwo
 \else \expandafter \@secondoftwo
 \fi
}%
\providecommand \@ifx [1]{%
 \ifx #1\expandafter \@firstoftwo
 \else \expandafter \@secondoftwo
 \fi
}%
\providecommand \natexlab [1]{#1}%
\providecommand \enquote  [1]{``#1''}%
\providecommand \bibnamefont  [1]{#1}%
\providecommand \bibfnamefont [1]{#1}%
\providecommand \citenamefont [1]{#1}%
\providecommand \href@noop [0]{\@secondoftwo}%
\providecommand \href [0]{\begingroup \@sanitize@url \@href}%
\providecommand \@href[1]{\@@startlink{#1}\@@href}%
\providecommand \@@href[1]{\endgroup#1\@@endlink}%
\providecommand \@sanitize@url [0]{\catcode `\\12\catcode `\$12\catcode
  `\&12\catcode `\#12\catcode `\^12\catcode `\_12\catcode `\%12\relax}%
\providecommand \@@startlink[1]{}%
\providecommand \@@endlink[0]{}%
\providecommand \url  [0]{\begingroup\@sanitize@url \@url }%
\providecommand \@url [1]{\endgroup\@href {#1}{\urlprefix }}%
\providecommand \urlprefix  [0]{URL }%
\providecommand \Eprint [0]{\href }%
\providecommand \doibase [0]{http://dx.doi.org/}%
\providecommand \selectlanguage [0]{\@gobble}%
\providecommand \bibinfo  [0]{\@secondoftwo}%
\providecommand \bibfield  [0]{\@secondoftwo}%
\providecommand \translation [1]{[#1]}%
\providecommand \BibitemOpen [0]{}%
\providecommand \bibitemStop [0]{}%
\providecommand \bibitemNoStop [0]{.\EOS\space}%
\providecommand \EOS [0]{\spacefactor3000\relax}%
\providecommand \BibitemShut  [1]{\csname bibitem#1\endcsname}%
\let\auto@bib@innerbib\@empty
%</preamble>
\bibitem [{\citenamefont {Bender}\ and\ \citenamefont
  {Boettcher}(1998)}]{bender1998real}%
  \BibitemOpen
  \bibfield  {author} {\bibinfo {author} {\bibfnamefont {C.~M.}\ \bibnamefont
  {Bender}}\ and\ \bibinfo {author} {\bibfnamefont {S.}~\bibnamefont
  {Boettcher}},\ }\href {\doibase 10.1103/PhysRevLett.80.5243} {\bibfield
  {journal} {\bibinfo  {journal} {Phys. Rev. Lett.}\ }\textbf {\bibinfo
  {volume} {80}},\ \bibinfo {pages} {5243} (\bibinfo {year}
  {1998})}\BibitemShut {NoStop}%
\bibitem [{\citenamefont {\mbox{C. M. Bender and
  company}}(2019)}]{bender2019pt}%
  \BibitemOpen
  \bibfield  {author} {\bibinfo {author} {\bibnamefont {\mbox{C. M. Bender and
  company}}},\ }\href@noop {} {\emph {\bibinfo {title} {PT Symmetry in Quantum
  and Classical Physics}}}\ (\bibinfo  {publisher} {World Scientific Publishing
  Europe},\ \bibinfo {year} {2019})\BibitemShut {NoStop}%
\bibitem [{\citenamefont {Bender}\ \emph {et~al.}(1999)\citenamefont {Bender},
  \citenamefont {Boettcher},\ and\ \citenamefont {Meisinger}}]{bender1999pt}%
  \BibitemOpen
  \bibfield  {author} {\bibinfo {author} {\bibfnamefont {C.~M.}\ \bibnamefont
  {Bender}}, \bibinfo {author} {\bibfnamefont {S.}~\bibnamefont {Boettcher}}, \
  and\ \bibinfo {author} {\bibfnamefont {P.~N.}\ \bibnamefont {Meisinger}},\
  }\href@noop {} {\bibfield  {journal} {\bibinfo  {journal} {Journal of
  Mathematical Physics}\ }\textbf {\bibinfo {volume} {40}},\ \bibinfo {pages}
  {2201} (\bibinfo {year} {1999})}\BibitemShut {NoStop}%
\bibitem [{\citenamefont {Bender}\ \emph {et~al.}(2002)\citenamefont {Bender},
  \citenamefont {Brody},\ and\ \citenamefont {Jones}}]{bender2002complex}%
  \BibitemOpen
  \bibfield  {author} {\bibinfo {author} {\bibfnamefont {C.~M.}\ \bibnamefont
  {Bender}}, \bibinfo {author} {\bibfnamefont {D.~C.}\ \bibnamefont {Brody}}, \
  and\ \bibinfo {author} {\bibfnamefont {H.~F.}\ \bibnamefont {Jones}},\ }\href
  {\doibase 10.1103/PhysRevLett.89.270401} {\bibfield  {journal} {\bibinfo
  {journal} {Phys. Rev. Lett.}\ }\textbf {\bibinfo {volume} {89}},\ \bibinfo
  {pages} {270401} (\bibinfo {year} {2002})}\BibitemShut {NoStop}%
\bibitem [{\citenamefont
  {Mostafazadeh}(2002{\natexlab{a}})}]{mostafazadeh2002pseudoI}%
  \BibitemOpen
  \bibfield  {author} {\bibinfo {author} {\bibfnamefont {A.}~\bibnamefont
  {Mostafazadeh}},\ }\href@noop {} {\bibfield  {journal} {\bibinfo  {journal}
  {Journal of Mathematical Physics}\ }\textbf {\bibinfo {volume} {43}},\
  \bibinfo {pages} {205} (\bibinfo {year} {2002}{\natexlab{a}})}\BibitemShut
  {NoStop}%
\bibitem [{\citenamefont
  {Mostafazadeh}(2002{\natexlab{b}})}]{mostafazadeh2002pseudoII}%
  \BibitemOpen
  \bibfield  {author} {\bibinfo {author} {\bibfnamefont {A.}~\bibnamefont
  {Mostafazadeh}},\ }\href@noop {} {\bibfield  {journal} {\bibinfo  {journal}
  {Journal of Mathematical Physics}\ }\textbf {\bibinfo {volume} {43}},\
  \bibinfo {pages} {2814} (\bibinfo {year} {2002}{\natexlab{b}})}\BibitemShut
  {NoStop}%
\bibitem [{\citenamefont
  {Mostafazadeh}(2002{\natexlab{c}})}]{mostafazadeh2002pseudoIII}%
  \BibitemOpen
  \bibfield  {author} {\bibinfo {author} {\bibfnamefont {A.}~\bibnamefont
  {Mostafazadeh}},\ }\href@noop {} {\bibfield  {journal} {\bibinfo  {journal}
  {Journal of Mathematical Physics}\ }\textbf {\bibinfo {volume} {43}},\
  \bibinfo {pages} {3944} (\bibinfo {year} {2002}{\natexlab{c}})}\BibitemShut
  {NoStop}%
\bibitem [{\citenamefont {Mostafazadeh}(2004)}]{mostafazadeh2004pseudounitary}%
  \BibitemOpen
  \bibfield  {author} {\bibinfo {author} {\bibfnamefont {A.}~\bibnamefont
  {Mostafazadeh}},\ }\href@noop {} {\bibfield  {journal} {\bibinfo  {journal}
  {Journal of mathematical physics}\ }\textbf {\bibinfo {volume} {45}},\
  \bibinfo {pages} {932} (\bibinfo {year} {2004})}\BibitemShut {NoStop}%
\bibitem [{\citenamefont {Bender}(2007)}]{bender2007making}%
  \BibitemOpen
  \bibfield  {author} {\bibinfo {author} {\bibfnamefont {C.~M.}\ \bibnamefont
  {Bender}},\ }\href@noop {} {\bibfield  {journal} {\bibinfo  {journal}
  {Reports on Progress in Physics}\ }\textbf {\bibinfo {volume} {70}},\
  \bibinfo {pages} {947} (\bibinfo {year} {2007})}\BibitemShut {NoStop}%
\bibitem [{\citenamefont {Brody}(2013)}]{brody2013biorthogonal}%
  \BibitemOpen
  \bibfield  {author} {\bibinfo {author} {\bibfnamefont {D.~C.}\ \bibnamefont
  {Brody}},\ }\href@noop {} {\bibfield  {journal} {\bibinfo  {journal} {Journal
  of Physics A: Mathematical and Theoretical}\ }\textbf {\bibinfo {volume}
  {47}},\ \bibinfo {pages} {035305} (\bibinfo {year} {2013})}\BibitemShut
  {NoStop}%
\bibitem [{\citenamefont {Guo}\ \emph {et~al.}(2009)\citenamefont {Guo},
  \citenamefont {Salamo}, \citenamefont {Duchesne}, \citenamefont {Morandotti},
  \citenamefont {Volatier-Ravat}, \citenamefont {Aimez}, \citenamefont
  {Siviloglou},\ and\ \citenamefont {Christodoulides}}]{guo2009observation}%
  \BibitemOpen
  \bibfield  {author} {\bibinfo {author} {\bibfnamefont {A.}~\bibnamefont
  {Guo}}, \bibinfo {author} {\bibfnamefont {G.~J.}\ \bibnamefont {Salamo}},
  \bibinfo {author} {\bibfnamefont {D.}~\bibnamefont {Duchesne}}, \bibinfo
  {author} {\bibfnamefont {R.}~\bibnamefont {Morandotti}}, \bibinfo {author}
  {\bibfnamefont {M.}~\bibnamefont {Volatier-Ravat}}, \bibinfo {author}
  {\bibfnamefont {V.}~\bibnamefont {Aimez}}, \bibinfo {author} {\bibfnamefont
  {G.~A.}\ \bibnamefont {Siviloglou}}, \ and\ \bibinfo {author} {\bibfnamefont
  {D.~N.}\ \bibnamefont {Christodoulides}},\ }\href {\doibase
  10.1103/PhysRevLett.103.093902} {\bibfield  {journal} {\bibinfo  {journal}
  {Phys. Rev. Lett.}\ }\textbf {\bibinfo {volume} {103}},\ \bibinfo {pages}
  {093902} (\bibinfo {year} {2009})}\BibitemShut {NoStop}%
\bibitem [{\citenamefont {Zheng}\ \emph {et~al.}(2010)\citenamefont {Zheng},
  \citenamefont {Christodoulides}, \citenamefont {Fleischmann},\ and\
  \citenamefont {Kottos}}]{zheng2010pt}%
  \BibitemOpen
  \bibfield  {author} {\bibinfo {author} {\bibfnamefont {M.~C.}\ \bibnamefont
  {Zheng}}, \bibinfo {author} {\bibfnamefont {D.~N.}\ \bibnamefont
  {Christodoulides}}, \bibinfo {author} {\bibfnamefont {R.}~\bibnamefont
  {Fleischmann}}, \ and\ \bibinfo {author} {\bibfnamefont {T.}~\bibnamefont
  {Kottos}},\ }\href {\doibase 10.1103/PhysRevA.82.010103} {\bibfield
  {journal} {\bibinfo  {journal} {Phys. Rev. A}\ }\textbf {\bibinfo {volume}
  {82}},\ \bibinfo {pages} {010103(R)} (\bibinfo {year} {2010})}\BibitemShut
  {NoStop}%
\bibitem [{\citenamefont {R{\"u}ter}\ \emph {et~al.}(2010)\citenamefont
  {R{\"u}ter}, \citenamefont {Makris}, \citenamefont {El-Ganainy},
  \citenamefont {Christodoulides}, \citenamefont {Segev},\ and\ \citenamefont
  {Kip}}]{ruter2010observation}%
  \BibitemOpen
  \bibfield  {author} {\bibinfo {author} {\bibfnamefont {C.~E.}\ \bibnamefont
  {R{\"u}ter}}, \bibinfo {author} {\bibfnamefont {K.~G.}\ \bibnamefont
  {Makris}}, \bibinfo {author} {\bibfnamefont {R.}~\bibnamefont {El-Ganainy}},
  \bibinfo {author} {\bibfnamefont {D.~N.}\ \bibnamefont {Christodoulides}},
  \bibinfo {author} {\bibfnamefont {M.}~\bibnamefont {Segev}}, \ and\ \bibinfo
  {author} {\bibfnamefont {D.}~\bibnamefont {Kip}},\ }\href@noop {} {\bibfield
  {journal} {\bibinfo  {journal} {Nature physics}\ }\textbf {\bibinfo {volume}
  {6}},\ \bibinfo {pages} {192} (\bibinfo {year} {2010})}\BibitemShut {NoStop}%
\bibitem [{\citenamefont {Miri}\ \emph {et~al.}(2012)\citenamefont {Miri},
  \citenamefont {LiKamWa},\ and\ \citenamefont
  {Christodoulides}}]{miri2012large}%
  \BibitemOpen
  \bibfield  {author} {\bibinfo {author} {\bibfnamefont {M.-A.}\ \bibnamefont
  {Miri}}, \bibinfo {author} {\bibfnamefont {P.}~\bibnamefont {LiKamWa}}, \
  and\ \bibinfo {author} {\bibfnamefont {D.~N.}\ \bibnamefont
  {Christodoulides}},\ }\href@noop {} {\bibfield  {journal} {\bibinfo
  {journal} {Optics letters}\ }\textbf {\bibinfo {volume} {37}},\ \bibinfo
  {pages} {764} (\bibinfo {year} {2012})}\BibitemShut {NoStop}%
\bibitem [{\citenamefont {Regensburger}\ \emph {et~al.}(2012)\citenamefont
  {Regensburger}, \citenamefont {Bersch}, \citenamefont {Miri}, \citenamefont
  {Onishchukov}, \citenamefont {Christodoulides},\ and\ \citenamefont
  {Peschel}}]{regensburger2012parity}%
  \BibitemOpen
  \bibfield  {author} {\bibinfo {author} {\bibfnamefont {A.}~\bibnamefont
  {Regensburger}}, \bibinfo {author} {\bibfnamefont {C.}~\bibnamefont
  {Bersch}}, \bibinfo {author} {\bibfnamefont {M.-A.}\ \bibnamefont {Miri}},
  \bibinfo {author} {\bibfnamefont {G.}~\bibnamefont {Onishchukov}}, \bibinfo
  {author} {\bibfnamefont {D.~N.}\ \bibnamefont {Christodoulides}}, \ and\
  \bibinfo {author} {\bibfnamefont {U.}~\bibnamefont {Peschel}},\ }\href@noop
  {} {\bibfield  {journal} {\bibinfo  {journal} {Nature}\ }\textbf {\bibinfo
  {volume} {488}},\ \bibinfo {pages} {167} (\bibinfo {year}
  {2012})}\BibitemShut {NoStop}%
\bibitem [{\citenamefont {Chtchelkatchev}\ \emph {et~al.}(2012)\citenamefont
  {Chtchelkatchev}, \citenamefont {Golubov}, \citenamefont {Baturina},\ and\
  \citenamefont {Vinokur}}]{chtchelkatchev2012stimulation}%
  \BibitemOpen
  \bibfield  {author} {\bibinfo {author} {\bibfnamefont {N.~M.}\ \bibnamefont
  {Chtchelkatchev}}, \bibinfo {author} {\bibfnamefont {A.~A.}\ \bibnamefont
  {Golubov}}, \bibinfo {author} {\bibfnamefont {T.~I.}\ \bibnamefont
  {Baturina}}, \ and\ \bibinfo {author} {\bibfnamefont {V.~M.}\ \bibnamefont
  {Vinokur}},\ }\href {\doibase 10.1103/PhysRevLett.109.150405} {\bibfield
  {journal} {\bibinfo  {journal} {Phys. Rev. Lett.}\ }\textbf {\bibinfo
  {volume} {109}},\ \bibinfo {pages} {150405} (\bibinfo {year}
  {2012})}\BibitemShut {NoStop}%
\bibitem [{\citenamefont {Schomerus}(2013)}]{schomerus2013topologically}%
  \BibitemOpen
  \bibfield  {author} {\bibinfo {author} {\bibfnamefont {H.}~\bibnamefont
  {Schomerus}},\ }\href@noop {} {\bibfield  {journal} {\bibinfo  {journal}
  {Optics letters}\ }\textbf {\bibinfo {volume} {38}},\ \bibinfo {pages} {1912}
  (\bibinfo {year} {2013})}\BibitemShut {NoStop}%
\bibitem [{\citenamefont {Feng}\ \emph {et~al.}(2014)\citenamefont {Feng},
  \citenamefont {Wong}, \citenamefont {Ma}, \citenamefont {Wang},\ and\
  \citenamefont {Zhang}}]{feng2014single}%
  \BibitemOpen
  \bibfield  {author} {\bibinfo {author} {\bibfnamefont {L.}~\bibnamefont
  {Feng}}, \bibinfo {author} {\bibfnamefont {Z.~J.}\ \bibnamefont {Wong}},
  \bibinfo {author} {\bibfnamefont {R.-M.}\ \bibnamefont {Ma}}, \bibinfo
  {author} {\bibfnamefont {Y.}~\bibnamefont {Wang}}, \ and\ \bibinfo {author}
  {\bibfnamefont {X.}~\bibnamefont {Zhang}},\ }\href@noop {} {\bibfield
  {journal} {\bibinfo  {journal} {Science}\ }\textbf {\bibinfo {volume}
  {346}},\ \bibinfo {pages} {972} (\bibinfo {year} {2014})}\BibitemShut
  {NoStop}%
\bibitem [{\citenamefont {Hodaei}\ \emph {et~al.}(2014)\citenamefont {Hodaei},
  \citenamefont {Miri}, \citenamefont {Heinrich}, \citenamefont
  {Christodoulides},\ and\ \citenamefont {Khajavikhan}}]{hodaei2014parity}%
  \BibitemOpen
  \bibfield  {author} {\bibinfo {author} {\bibfnamefont {H.}~\bibnamefont
  {Hodaei}}, \bibinfo {author} {\bibfnamefont {M.-A.}\ \bibnamefont {Miri}},
  \bibinfo {author} {\bibfnamefont {M.}~\bibnamefont {Heinrich}}, \bibinfo
  {author} {\bibfnamefont {D.~N.}\ \bibnamefont {Christodoulides}}, \ and\
  \bibinfo {author} {\bibfnamefont {M.}~\bibnamefont {Khajavikhan}},\
  }\href@noop {} {\bibfield  {journal} {\bibinfo  {journal} {Science}\ }\textbf
  {\bibinfo {volume} {346}},\ \bibinfo {pages} {975} (\bibinfo {year}
  {2014})}\BibitemShut {NoStop}%
\bibitem [{\citenamefont {Peng}\ \emph
  {et~al.}(2014{\natexlab{a}})\citenamefont {Peng}, \citenamefont
  {{\"O}zdemir}, \citenamefont {Lei}, \citenamefont {Monifi}, \citenamefont
  {Gianfreda}, \citenamefont {Long}, \citenamefont {Fan}, \citenamefont {Nori},
  \citenamefont {Bender},\ and\ \citenamefont {Yang}}]{peng2014parity}%
  \BibitemOpen
  \bibfield  {author} {\bibinfo {author} {\bibfnamefont {B.}~\bibnamefont
  {Peng}}, \bibinfo {author} {\bibfnamefont {{\c{S}}.~K.}\ \bibnamefont
  {{\"O}zdemir}}, \bibinfo {author} {\bibfnamefont {F.}~\bibnamefont {Lei}},
  \bibinfo {author} {\bibfnamefont {F.}~\bibnamefont {Monifi}}, \bibinfo
  {author} {\bibfnamefont {M.}~\bibnamefont {Gianfreda}}, \bibinfo {author}
  {\bibfnamefont {G.~L.}\ \bibnamefont {Long}}, \bibinfo {author}
  {\bibfnamefont {S.}~\bibnamefont {Fan}}, \bibinfo {author} {\bibfnamefont
  {F.}~\bibnamefont {Nori}}, \bibinfo {author} {\bibfnamefont {C.~M.}\
  \bibnamefont {Bender}}, \ and\ \bibinfo {author} {\bibfnamefont
  {L.}~\bibnamefont {Yang}},\ }\href@noop {} {\bibfield  {journal} {\bibinfo
  {journal} {Nature Physics}\ }\textbf {\bibinfo {volume} {10}},\ \bibinfo
  {pages} {394} (\bibinfo {year} {2014}{\natexlab{a}})}\BibitemShut {NoStop}%
\bibitem [{\citenamefont {Peng}\ \emph
  {et~al.}(2014{\natexlab{b}})\citenamefont {Peng}, \citenamefont
  {{\"O}zdemir}, \citenamefont {Rotter}, \citenamefont {Yilmaz}, \citenamefont
  {Liertzer}, \citenamefont {Monifi}, \citenamefont {Bender}, \citenamefont
  {Nori},\ and\ \citenamefont {Yang}}]{peng2014loss}%
  \BibitemOpen
  \bibfield  {author} {\bibinfo {author} {\bibfnamefont {B.}~\bibnamefont
  {Peng}}, \bibinfo {author} {\bibfnamefont {{\c{S}}.}~\bibnamefont
  {{\"O}zdemir}}, \bibinfo {author} {\bibfnamefont {S.}~\bibnamefont {Rotter}},
  \bibinfo {author} {\bibfnamefont {H.}~\bibnamefont {Yilmaz}}, \bibinfo
  {author} {\bibfnamefont {M.}~\bibnamefont {Liertzer}}, \bibinfo {author}
  {\bibfnamefont {F.}~\bibnamefont {Monifi}}, \bibinfo {author} {\bibfnamefont
  {C.}~\bibnamefont {Bender}}, \bibinfo {author} {\bibfnamefont
  {F.}~\bibnamefont {Nori}}, \ and\ \bibinfo {author} {\bibfnamefont
  {L.}~\bibnamefont {Yang}},\ }\href@noop {} {\bibfield  {journal} {\bibinfo
  {journal} {Science}\ }\textbf {\bibinfo {volume} {346}},\ \bibinfo {pages}
  {328} (\bibinfo {year} {2014}{\natexlab{b}})}\BibitemShut {NoStop}%
\bibitem [{\citenamefont {Poli}\ \emph {et~al.}(2015)\citenamefont {Poli},
  \citenamefont {Bellec}, \citenamefont {Kuhl}, \citenamefont {Mortessagne},\
  and\ \citenamefont {Schomerus}}]{poli2015selective}%
  \BibitemOpen
  \bibfield  {author} {\bibinfo {author} {\bibfnamefont {C.}~\bibnamefont
  {Poli}}, \bibinfo {author} {\bibfnamefont {M.}~\bibnamefont {Bellec}},
  \bibinfo {author} {\bibfnamefont {U.}~\bibnamefont {Kuhl}}, \bibinfo {author}
  {\bibfnamefont {F.}~\bibnamefont {Mortessagne}}, \ and\ \bibinfo {author}
  {\bibfnamefont {H.}~\bibnamefont {Schomerus}},\ }\href@noop {} {\bibfield
  {journal} {\bibinfo  {journal} {Nature communications}\ }\textbf {\bibinfo
  {volume} {6}},\ \bibinfo {pages} {6710} (\bibinfo {year} {2015})}\BibitemShut
  {NoStop}%
\bibitem [{\citenamefont {Zeuner}\ \emph {et~al.}(2015)\citenamefont {Zeuner},
  \citenamefont {Rechtsman}, \citenamefont {Plotnik}, \citenamefont {Lumer},
  \citenamefont {Nolte}, \citenamefont {Rudner}, \citenamefont {Segev},\ and\
  \citenamefont {Szameit}}]{zeuner2015observation}%
  \BibitemOpen
  \bibfield  {author} {\bibinfo {author} {\bibfnamefont {J.~M.}\ \bibnamefont
  {Zeuner}}, \bibinfo {author} {\bibfnamefont {M.~C.}\ \bibnamefont
  {Rechtsman}}, \bibinfo {author} {\bibfnamefont {Y.}~\bibnamefont {Plotnik}},
  \bibinfo {author} {\bibfnamefont {Y.}~\bibnamefont {Lumer}}, \bibinfo
  {author} {\bibfnamefont {S.}~\bibnamefont {Nolte}}, \bibinfo {author}
  {\bibfnamefont {M.~S.}\ \bibnamefont {Rudner}}, \bibinfo {author}
  {\bibfnamefont {M.}~\bibnamefont {Segev}}, \ and\ \bibinfo {author}
  {\bibfnamefont {A.}~\bibnamefont {Szameit}},\ }\href {\doibase
  10.1103/PhysRevLett.115.040402} {\bibfield  {journal} {\bibinfo  {journal}
  {Phys. Rev. Lett.}\ }\textbf {\bibinfo {volume} {115}},\ \bibinfo {pages}
  {040402} (\bibinfo {year} {2015})}\BibitemShut {NoStop}%
\bibitem [{\citenamefont {Mochizuki}\ \emph {et~al.}(2016)\citenamefont
  {Mochizuki}, \citenamefont {Kim},\ and\ \citenamefont
  {Obuse}}]{mochizuki2016explicit}%
  \BibitemOpen
  \bibfield  {author} {\bibinfo {author} {\bibfnamefont {K.}~\bibnamefont
  {Mochizuki}}, \bibinfo {author} {\bibfnamefont {D.}~\bibnamefont {Kim}}, \
  and\ \bibinfo {author} {\bibfnamefont {H.}~\bibnamefont {Obuse}},\ }\href
  {\doibase 10.1103/PhysRevA.93.062116} {\bibfield  {journal} {\bibinfo
  {journal} {Phys. Rev. A}\ }\textbf {\bibinfo {volume} {93}},\ \bibinfo
  {pages} {062116} (\bibinfo {year} {2016})}\BibitemShut {NoStop}%
\bibitem [{\citenamefont {Kim}\ \emph {et~al.}(2016)\citenamefont {Kim},
  \citenamefont {Ken}, \citenamefont {Kawakami},\ and\ \citenamefont
  {Obuse}}]{kim2016floquet}%
  \BibitemOpen
  \bibfield  {author} {\bibinfo {author} {\bibfnamefont {D.}~\bibnamefont
  {Kim}}, \bibinfo {author} {\bibfnamefont {M.}~\bibnamefont {Ken}}, \bibinfo
  {author} {\bibfnamefont {N.}~\bibnamefont {Kawakami}}, \ and\ \bibinfo
  {author} {\bibfnamefont {H.}~\bibnamefont {Obuse}},\ }\href@noop {}
  {\bibfield  {journal} {\bibinfo  {journal} {arXiv preprint arXiv:1609.09650}\
  } (\bibinfo {year} {2016})}\BibitemShut {NoStop}%
\bibitem [{\citenamefont {Xiao}\ \emph {et~al.}(2017)\citenamefont {Xiao},
  \citenamefont {Zhan}, \citenamefont {Bian}, \citenamefont {Wang},
  \citenamefont {Zhang}, \citenamefont {Wang}, \citenamefont {Li},
  \citenamefont {Mochizuki}, \citenamefont {Kim}, \citenamefont {Kawakami},
  \citenamefont {Yi}, \citenamefont {Obuse}, \citenamefont {Sanders},\ and\
  \citenamefont {Xue}}]{xiao2017observation}%
  \BibitemOpen
  \bibfield  {author} {\bibinfo {author} {\bibfnamefont {L.}~\bibnamefont
  {Xiao}}, \bibinfo {author} {\bibfnamefont {X.}~\bibnamefont {Zhan}}, \bibinfo
  {author} {\bibfnamefont {Z.}~\bibnamefont {Bian}}, \bibinfo {author}
  {\bibfnamefont {K.}~\bibnamefont {Wang}}, \bibinfo {author} {\bibfnamefont
  {X.}~\bibnamefont {Zhang}}, \bibinfo {author} {\bibfnamefont
  {X.}~\bibnamefont {Wang}}, \bibinfo {author} {\bibfnamefont {J.}~\bibnamefont
  {Li}}, \bibinfo {author} {\bibfnamefont {K.}~\bibnamefont {Mochizuki}},
  \bibinfo {author} {\bibfnamefont {D.}~\bibnamefont {Kim}}, \bibinfo {author}
  {\bibfnamefont {N.}~\bibnamefont {Kawakami}}, \bibinfo {author}
  {\bibfnamefont {W.}~\bibnamefont {Yi}}, \bibinfo {author} {\bibfnamefont
  {H.}~\bibnamefont {Obuse}}, \bibinfo {author} {\bibfnamefont {B.~C.}\
  \bibnamefont {Sanders}}, \ and\ \bibinfo {author} {\bibfnamefont
  {P.}~\bibnamefont {Xue}},\ }\href@noop {} {\bibfield  {journal} {\bibinfo
  {journal} {Nature Physics}\ }\textbf {\bibinfo {volume} {13}},\ \bibinfo
  {pages} {1117} (\bibinfo {year} {2017})}\BibitemShut {NoStop}%
\bibitem [{\citenamefont {Hu}\ and\ \citenamefont
  {Hughes}(2011)}]{hu2011absence}%
  \BibitemOpen
  \bibfield  {author} {\bibinfo {author} {\bibfnamefont {Y.~C.}\ \bibnamefont
  {Hu}}\ and\ \bibinfo {author} {\bibfnamefont {T.~L.}\ \bibnamefont
  {Hughes}},\ }\href {\doibase 10.1103/PhysRevB.84.153101} {\bibfield
  {journal} {\bibinfo  {journal} {Phys. Rev. B}\ }\textbf {\bibinfo {volume}
  {84}},\ \bibinfo {pages} {153101} (\bibinfo {year} {2011})}\BibitemShut
  {NoStop}%
\bibitem [{\citenamefont {Esaki}\ \emph {et~al.}(2011)\citenamefont {Esaki},
  \citenamefont {Sato}, \citenamefont {Hasebe},\ and\ \citenamefont
  {Kohmoto}}]{esaki2011edge}%
  \BibitemOpen
  \bibfield  {author} {\bibinfo {author} {\bibfnamefont {K.}~\bibnamefont
  {Esaki}}, \bibinfo {author} {\bibfnamefont {M.}~\bibnamefont {Sato}},
  \bibinfo {author} {\bibfnamefont {K.}~\bibnamefont {Hasebe}}, \ and\ \bibinfo
  {author} {\bibfnamefont {M.}~\bibnamefont {Kohmoto}},\ }\href {\doibase
  10.1103/PhysRevB.84.205128} {\bibfield  {journal} {\bibinfo  {journal} {Phys.
  Rev. B}\ }\textbf {\bibinfo {volume} {84}},\ \bibinfo {pages} {205128}
  (\bibinfo {year} {2011})}\BibitemShut {NoStop}%
\bibitem [{\citenamefont {Leykam}\ \emph {et~al.}(2017)\citenamefont {Leykam},
  \citenamefont {Bliokh}, \citenamefont {Huang}, \citenamefont {Chong},\ and\
  \citenamefont {Nori}}]{leykam2017edge}%
  \BibitemOpen
  \bibfield  {author} {\bibinfo {author} {\bibfnamefont {D.}~\bibnamefont
  {Leykam}}, \bibinfo {author} {\bibfnamefont {K.~Y.}\ \bibnamefont {Bliokh}},
  \bibinfo {author} {\bibfnamefont {C.}~\bibnamefont {Huang}}, \bibinfo
  {author} {\bibfnamefont {Y.~D.}\ \bibnamefont {Chong}}, \ and\ \bibinfo
  {author} {\bibfnamefont {F.}~\bibnamefont {Nori}},\ }\href {\doibase
  10.1103/PhysRevLett.118.040401} {\bibfield  {journal} {\bibinfo  {journal}
  {Phys. Rev. Lett.}\ }\textbf {\bibinfo {volume} {118}},\ \bibinfo {pages}
  {040401} (\bibinfo {year} {2017})}\BibitemShut {NoStop}%
\bibitem [{\citenamefont {Martinez~Alvarez}\ \emph {et~al.}(2018)\citenamefont
  {Martinez~Alvarez}, \citenamefont {Barrios~Vargas},\ and\ \citenamefont
  {Foa~Torres}}]{alvarez2018non}%
  \BibitemOpen
  \bibfield  {author} {\bibinfo {author} {\bibfnamefont {V.~M.}\ \bibnamefont
  {Martinez~Alvarez}}, \bibinfo {author} {\bibfnamefont {J.~E.}\ \bibnamefont
  {Barrios~Vargas}}, \ and\ \bibinfo {author} {\bibfnamefont {L.~E.~F.}\
  \bibnamefont {Foa~Torres}},\ }\href {\doibase 10.1103/PhysRevB.97.121401}
  {\bibfield  {journal} {\bibinfo  {journal} {Phys. Rev. B}\ }\textbf {\bibinfo
  {volume} {97}},\ \bibinfo {pages} {121401(R)} (\bibinfo {year}
  {2018})}\BibitemShut {NoStop}%
\bibitem [{\citenamefont {Shen}\ \emph {et~al.}(2018)\citenamefont {Shen},
  \citenamefont {Zhen},\ and\ \citenamefont {Fu}}]{shen2018topological}%
  \BibitemOpen
  \bibfield  {author} {\bibinfo {author} {\bibfnamefont {H.}~\bibnamefont
  {Shen}}, \bibinfo {author} {\bibfnamefont {B.}~\bibnamefont {Zhen}}, \ and\
  \bibinfo {author} {\bibfnamefont {L.}~\bibnamefont {Fu}},\ }\href {\doibase
  10.1103/PhysRevLett.120.146402} {\bibfield  {journal} {\bibinfo  {journal}
  {Phys. Rev. Lett.}\ }\textbf {\bibinfo {volume} {120}},\ \bibinfo {pages}
  {146402} (\bibinfo {year} {2018})}\BibitemShut {NoStop}%
\bibitem [{\citenamefont {Rudner}\ and\ \citenamefont
  {Levitov}(2009)}]{rudner2009topological}%
  \BibitemOpen
  \bibfield  {author} {\bibinfo {author} {\bibfnamefont {M.~S.}\ \bibnamefont
  {Rudner}}\ and\ \bibinfo {author} {\bibfnamefont {L.~S.}\ \bibnamefont
  {Levitov}},\ }\href {\doibase 10.1103/PhysRevLett.102.065703} {\bibfield
  {journal} {\bibinfo  {journal} {Phys. Rev. Lett.}\ }\textbf {\bibinfo
  {volume} {102}},\ \bibinfo {pages} {065703} (\bibinfo {year}
  {2009})}\BibitemShut {NoStop}%
\bibitem [{\citenamefont {Weimann}\ \emph {et~al.}(2017)\citenamefont
  {Weimann}, \citenamefont {Kremer}, \citenamefont {Plotnik}, \citenamefont
  {Lumer}, \citenamefont {Nolte}, \citenamefont {Makris}, \citenamefont
  {Segev}, \citenamefont {Rechtsman},\ and\ \citenamefont
  {Szameit}}]{weimann2017topologically}%
  \BibitemOpen
  \bibfield  {author} {\bibinfo {author} {\bibfnamefont {S.}~\bibnamefont
  {Weimann}}, \bibinfo {author} {\bibfnamefont {M.}~\bibnamefont {Kremer}},
  \bibinfo {author} {\bibfnamefont {Y.}~\bibnamefont {Plotnik}}, \bibinfo
  {author} {\bibfnamefont {Y.}~\bibnamefont {Lumer}}, \bibinfo {author}
  {\bibfnamefont {S.}~\bibnamefont {Nolte}}, \bibinfo {author} {\bibfnamefont
  {K.~G.}\ \bibnamefont {Makris}}, \bibinfo {author} {\bibfnamefont
  {M.}~\bibnamefont {Segev}}, \bibinfo {author} {\bibfnamefont {M.~C.}\
  \bibnamefont {Rechtsman}}, \ and\ \bibinfo {author} {\bibfnamefont
  {A.}~\bibnamefont {Szameit}},\ }\href@noop {} {\bibfield  {journal} {\bibinfo
   {journal} {Nature materials}\ }\textbf {\bibinfo {volume} {16}},\ \bibinfo
  {pages} {433} (\bibinfo {year} {2017})}\BibitemShut {NoStop}%
\bibitem [{\citenamefont {Kunst}\ \emph {et~al.}(2018)\citenamefont {Kunst},
  \citenamefont {Edvardsson}, \citenamefont {Budich},\ and\ \citenamefont
  {Bergholtz}}]{kunst2018biorthogonal}%
  \BibitemOpen
  \bibfield  {author} {\bibinfo {author} {\bibfnamefont {F.~K.}\ \bibnamefont
  {Kunst}}, \bibinfo {author} {\bibfnamefont {E.}~\bibnamefont {Edvardsson}},
  \bibinfo {author} {\bibfnamefont {J.~C.}\ \bibnamefont {Budich}}, \ and\
  \bibinfo {author} {\bibfnamefont {E.~J.}\ \bibnamefont {Bergholtz}},\ }\href
  {\doibase 10.1103/PhysRevLett.121.026808} {\bibfield  {journal} {\bibinfo
  {journal} {Phys. Rev. Lett.}\ }\textbf {\bibinfo {volume} {121}},\ \bibinfo
  {pages} {026808} (\bibinfo {year} {2018})}\BibitemShut {NoStop}%
\bibitem [{\citenamefont {Qi}\ \emph {et~al.}(2018)\citenamefont {Qi},
  \citenamefont {Zhang},\ and\ \citenamefont {Ge}}]{qi2018defect}%
  \BibitemOpen
  \bibfield  {author} {\bibinfo {author} {\bibfnamefont {B.}~\bibnamefont
  {Qi}}, \bibinfo {author} {\bibfnamefont {L.}~\bibnamefont {Zhang}}, \ and\
  \bibinfo {author} {\bibfnamefont {L.}~\bibnamefont {Ge}},\ }\href {\doibase
  10.1103/PhysRevLett.120.093901} {\bibfield  {journal} {\bibinfo  {journal}
  {Phys. Rev. Lett.}\ }\textbf {\bibinfo {volume} {120}},\ \bibinfo {pages}
  {093901} (\bibinfo {year} {2018})}\BibitemShut {NoStop}%
\bibitem [{\citenamefont {Lieu}(2018)}]{lieu2018topological}%
  \BibitemOpen
  \bibfield  {author} {\bibinfo {author} {\bibfnamefont {S.}~\bibnamefont
  {Lieu}},\ }\href {\doibase 10.1103/PhysRevB.97.045106} {\bibfield  {journal}
  {\bibinfo  {journal} {Phys. Rev. B}\ }\textbf {\bibinfo {volume} {97}},\
  \bibinfo {pages} {045106} (\bibinfo {year} {2018})}\BibitemShut {NoStop}%
\bibitem [{\citenamefont {Dangel}\ \emph {et~al.}(2018)\citenamefont {Dangel},
  \citenamefont {Wagner}, \citenamefont {Cartarius}, \citenamefont {Main},\
  and\ \citenamefont {Wunner}}]{dangel2018topological}%
  \BibitemOpen
  \bibfield  {author} {\bibinfo {author} {\bibfnamefont {F.}~\bibnamefont
  {Dangel}}, \bibinfo {author} {\bibfnamefont {M.}~\bibnamefont {Wagner}},
  \bibinfo {author} {\bibfnamefont {H.}~\bibnamefont {Cartarius}}, \bibinfo
  {author} {\bibfnamefont {J.}~\bibnamefont {Main}}, \ and\ \bibinfo {author}
  {\bibfnamefont {G.}~\bibnamefont {Wunner}},\ }\href {\doibase
  10.1103/PhysRevA.98.013628} {\bibfield  {journal} {\bibinfo  {journal} {Phys.
  Rev. A}\ }\textbf {\bibinfo {volume} {98}},\ \bibinfo {pages} {013628}
  (\bibinfo {year} {2018})}\BibitemShut {NoStop}%
\bibitem [{\citenamefont {Yao}\ and\ \citenamefont {Wang}(2018)}]{yao2018edge}%
  \BibitemOpen
  \bibfield  {author} {\bibinfo {author} {\bibfnamefont {S.}~\bibnamefont
  {Yao}}\ and\ \bibinfo {author} {\bibfnamefont {Z.}~\bibnamefont {Wang}},\
  }\href {\doibase 10.1103/PhysRevLett.121.086803} {\bibfield  {journal}
  {\bibinfo  {journal} {Phys. Rev. Lett.}\ }\textbf {\bibinfo {volume} {121}},\
  \bibinfo {pages} {086803} (\bibinfo {year} {2018})}\BibitemShut {NoStop}%
\bibitem [{\citenamefont {Ghatak}\ and\ \citenamefont
  {Das}(2019)}]{ghatak2019new}%
  \BibitemOpen
  \bibfield  {author} {\bibinfo {author} {\bibfnamefont {A.}~\bibnamefont
  {Ghatak}}\ and\ \bibinfo {author} {\bibfnamefont {T.}~\bibnamefont {Das}},\
  }\href@noop {} {\bibfield  {journal} {\bibinfo  {journal} {Journal of
  Physics: Condensed Matter}\ }\textbf {\bibinfo {volume} {31}},\ \bibinfo
  {pages} {263001} (\bibinfo {year} {2019})}\BibitemShut {NoStop}%
\bibitem [{\citenamefont {Kawabata}\ \emph {et~al.}(2019)\citenamefont
  {Kawabata}, \citenamefont {Shiozaki}, \citenamefont {Ueda},\ and\
  \citenamefont {Sato}}]{kawabata2019symmetry}%
  \BibitemOpen
  \bibfield  {author} {\bibinfo {author} {\bibfnamefont {K.}~\bibnamefont
  {Kawabata}}, \bibinfo {author} {\bibfnamefont {K.}~\bibnamefont {Shiozaki}},
  \bibinfo {author} {\bibfnamefont {M.}~\bibnamefont {Ueda}}, \ and\ \bibinfo
  {author} {\bibfnamefont {M.}~\bibnamefont {Sato}},\ }\href {\doibase
  10.1103/PhysRevX.9.041015} {\bibfield  {journal} {\bibinfo  {journal} {Phys.
  Rev. X}\ }\textbf {\bibinfo {volume} {9}},\ \bibinfo {pages} {041015}
  (\bibinfo {year} {2019})}\BibitemShut {NoStop}%
\bibitem [{\citenamefont {Borgnia}\ \emph {et~al.}(2020)\citenamefont
  {Borgnia}, \citenamefont {Kruchkov},\ and\ \citenamefont
  {Slager}}]{borgnia2020non}%
  \BibitemOpen
  \bibfield  {author} {\bibinfo {author} {\bibfnamefont {D.~S.}\ \bibnamefont
  {Borgnia}}, \bibinfo {author} {\bibfnamefont {A.~J.}\ \bibnamefont
  {Kruchkov}}, \ and\ \bibinfo {author} {\bibfnamefont {R.-J.}\ \bibnamefont
  {Slager}},\ }\href {\doibase 10.1103/PhysRevLett.124.056802} {\bibfield
  {journal} {\bibinfo  {journal} {Phys. Rev. Lett.}\ }\textbf {\bibinfo
  {volume} {124}},\ \bibinfo {pages} {056802} (\bibinfo {year}
  {2020})}\BibitemShut {NoStop}%
\bibitem [{\citenamefont {Xiao}\ \emph {et~al.}(2020)\citenamefont {Xiao},
  \citenamefont {Deng}, \citenamefont {Wang}, \citenamefont {Zhu},
  \citenamefont {Wang}, \citenamefont {Yi},\ and\ \citenamefont
  {Xue}}]{xiao2020non}%
  \BibitemOpen
  \bibfield  {author} {\bibinfo {author} {\bibfnamefont {L.}~\bibnamefont
  {Xiao}}, \bibinfo {author} {\bibfnamefont {T.}~\bibnamefont {Deng}}, \bibinfo
  {author} {\bibfnamefont {K.}~\bibnamefont {Wang}}, \bibinfo {author}
  {\bibfnamefont {G.}~\bibnamefont {Zhu}}, \bibinfo {author} {\bibfnamefont
  {Z.}~\bibnamefont {Wang}}, \bibinfo {author} {\bibfnamefont {W.}~\bibnamefont
  {Yi}}, \ and\ \bibinfo {author} {\bibfnamefont {P.}~\bibnamefont {Xue}},\
  }\href {\doibase 10.1038/s41567-020-0836-6} {\bibfield  {journal} {\bibinfo
  {journal} {Nature Physics}\ ,\ \bibinfo {pages} {1}} (\bibinfo {year}
  {2020})}\BibitemShut {NoStop}%
\bibitem [{\citenamefont {Hatano}\ and\ \citenamefont
  {Nelson}(1996)}]{hatano1996localization}%
  \BibitemOpen
  \bibfield  {author} {\bibinfo {author} {\bibfnamefont {N.}~\bibnamefont
  {Hatano}}\ and\ \bibinfo {author} {\bibfnamefont {D.~R.}\ \bibnamefont
  {Nelson}},\ }\href {\doibase 10.1103/PhysRevLett.77.570} {\bibfield
  {journal} {\bibinfo  {journal} {Phys. Rev. Lett.}\ }\textbf {\bibinfo
  {volume} {77}},\ \bibinfo {pages} {570} (\bibinfo {year} {1996})}\BibitemShut
  {NoStop}%
\bibitem [{\citenamefont {Hatano}\ and\ \citenamefont
  {Nelson}(1997)}]{hatano1997vortex}%
  \BibitemOpen
  \bibfield  {author} {\bibinfo {author} {\bibfnamefont {N.}~\bibnamefont
  {Hatano}}\ and\ \bibinfo {author} {\bibfnamefont {D.~R.}\ \bibnamefont
  {Nelson}},\ }\href {\doibase 10.1103/PhysRevB.56.8651} {\bibfield  {journal}
  {\bibinfo  {journal} {Phys. Rev. B}\ }\textbf {\bibinfo {volume} {56}},\
  \bibinfo {pages} {8651} (\bibinfo {year} {1997})}\BibitemShut {NoStop}%
\bibitem [{\citenamefont {Hatano}\ and\ \citenamefont
  {Nelson}(1998)}]{hatano1998non}%
  \BibitemOpen
  \bibfield  {author} {\bibinfo {author} {\bibfnamefont {N.}~\bibnamefont
  {Hatano}}\ and\ \bibinfo {author} {\bibfnamefont {D.~R.}\ \bibnamefont
  {Nelson}},\ }\href {\doibase 10.1103/PhysRevB.58.8384} {\bibfield  {journal}
  {\bibinfo  {journal} {Phys. Rev. B}\ }\textbf {\bibinfo {volume} {58}},\
  \bibinfo {pages} {8384} (\bibinfo {year} {1998})}\BibitemShut {NoStop}%
\bibitem [{\citenamefont {Goldsheid}\ and\ \citenamefont
  {Khoruzhenko}(1998)}]{goldsheid1998distribution}%
  \BibitemOpen
  \bibfield  {author} {\bibinfo {author} {\bibfnamefont {I.~Y.}\ \bibnamefont
  {Goldsheid}}\ and\ \bibinfo {author} {\bibfnamefont {B.~A.}\ \bibnamefont
  {Khoruzhenko}},\ }\href {\doibase 10.1103/PhysRevLett.80.2897} {\bibfield
  {journal} {\bibinfo  {journal} {Phys. Rev. Lett.}\ }\textbf {\bibinfo
  {volume} {80}},\ \bibinfo {pages} {2897} (\bibinfo {year}
  {1998})}\BibitemShut {NoStop}%
\bibitem [{\citenamefont {Shnerb}\ and\ \citenamefont
  {Nelson}(1998)}]{shnerb1998winding}%
  \BibitemOpen
  \bibfield  {author} {\bibinfo {author} {\bibfnamefont {N.~M.}\ \bibnamefont
  {Shnerb}}\ and\ \bibinfo {author} {\bibfnamefont {D.~R.}\ \bibnamefont
  {Nelson}},\ }\href {\doibase 10.1103/PhysRevLett.80.5172} {\bibfield
  {journal} {\bibinfo  {journal} {Phys. Rev. Lett.}\ }\textbf {\bibinfo
  {volume} {80}},\ \bibinfo {pages} {5172} (\bibinfo {year}
  {1998})}\BibitemShut {NoStop}%
\bibitem [{\citenamefont {Feinberg}\ and\ \citenamefont
  {Zee}(1999{\natexlab{a}})}]{feinberg1999non}%
  \BibitemOpen
  \bibfield  {author} {\bibinfo {author} {\bibfnamefont {J.}~\bibnamefont
  {Feinberg}}\ and\ \bibinfo {author} {\bibfnamefont {A.}~\bibnamefont {Zee}},\
  }\href {\doibase 10.1103/PhysRevE.59.6433} {\bibfield  {journal} {\bibinfo
  {journal} {Phys. Rev. E}\ }\textbf {\bibinfo {volume} {59}},\ \bibinfo
  {pages} {6433} (\bibinfo {year} {1999}{\natexlab{a}})}\BibitemShut {NoStop}%
\bibitem [{\citenamefont {Feinberg}\ and\ \citenamefont
  {Zee}(1999{\natexlab{b}})}]{feinberg1999curves}%
  \BibitemOpen
  \bibfield  {author} {\bibinfo {author} {\bibfnamefont {J.}~\bibnamefont
  {Feinberg}}\ and\ \bibinfo {author} {\bibfnamefont {A.}~\bibnamefont {Zee}},\
  }\href {\doibase 10.1016/S0550-3213(99)00246-1} {\bibfield  {journal}
  {\bibinfo  {journal} {Nucl. Phys. B}\ }\textbf {\bibinfo {volume} {552}},\
  \bibinfo {pages} {599} (\bibinfo {year} {1999}{\natexlab{b}})}\BibitemShut
  {NoStop}%
\bibitem [{\citenamefont {Kolesnikov}\ and\ \citenamefont
  {Efetov}(2000)}]{kolesnikov2000localization}%
  \BibitemOpen
  \bibfield  {author} {\bibinfo {author} {\bibfnamefont {A.~V.}\ \bibnamefont
  {Kolesnikov}}\ and\ \bibinfo {author} {\bibfnamefont {K.~B.}\ \bibnamefont
  {Efetov}},\ }\href {\doibase 10.1103/PhysRevLett.84.5600} {\bibfield
  {journal} {\bibinfo  {journal} {Phys. Rev. Lett.}\ }\textbf {\bibinfo
  {volume} {84}},\ \bibinfo {pages} {5600} (\bibinfo {year}
  {2000})}\BibitemShut {NoStop}%
\bibitem [{\citenamefont {Moiseyev}\ and\ \citenamefont
  {Gl\"uck}(2001)}]{moiseyev2001non}%
  \BibitemOpen
  \bibfield  {author} {\bibinfo {author} {\bibfnamefont {N.}~\bibnamefont
  {Moiseyev}}\ and\ \bibinfo {author} {\bibfnamefont {M.}~\bibnamefont
  {Gl\"uck}},\ }\href {\doibase 10.1103/PhysRevE.63.041103} {\bibfield
  {journal} {\bibinfo  {journal} {Phys. Rev. E}\ }\textbf {\bibinfo {volume}
  {63}},\ \bibinfo {pages} {041103} (\bibinfo {year} {2001})}\BibitemShut
  {NoStop}%
\bibitem [{\citenamefont {Amir}\ \emph {et~al.}(2016)\citenamefont {Amir},
  \citenamefont {Hatano},\ and\ \citenamefont {Nelson}}]{amir2016non}%
  \BibitemOpen
  \bibfield  {author} {\bibinfo {author} {\bibfnamefont {A.}~\bibnamefont
  {Amir}}, \bibinfo {author} {\bibfnamefont {N.}~\bibnamefont {Hatano}}, \ and\
  \bibinfo {author} {\bibfnamefont {D.~R.}\ \bibnamefont {Nelson}},\ }\href
  {\doibase 10.1103/PhysRevE.93.042310} {\bibfield  {journal} {\bibinfo
  {journal} {Phys. Rev. E}\ }\textbf {\bibinfo {volume} {93}},\ \bibinfo
  {pages} {042310} (\bibinfo {year} {2016})}\BibitemShut {NoStop}%
\bibitem [{\citenamefont {Jiang}\ \emph {et~al.}(2019)\citenamefont {Jiang},
  \citenamefont {Lang}, \citenamefont {Yang}, \citenamefont {Zhu},\ and\
  \citenamefont {Chen}}]{jiang2019interplay}%
  \BibitemOpen
  \bibfield  {author} {\bibinfo {author} {\bibfnamefont {H.}~\bibnamefont
  {Jiang}}, \bibinfo {author} {\bibfnamefont {L.-J.}\ \bibnamefont {Lang}},
  \bibinfo {author} {\bibfnamefont {C.}~\bibnamefont {Yang}}, \bibinfo {author}
  {\bibfnamefont {S.-L.}\ \bibnamefont {Zhu}}, \ and\ \bibinfo {author}
  {\bibfnamefont {S.}~\bibnamefont {Chen}},\ }\href {\doibase
  10.1103/PhysRevB.100.054301} {\bibfield  {journal} {\bibinfo  {journal}
  {Phys. Rev. B}\ }\textbf {\bibinfo {volume} {100}},\ \bibinfo {pages}
  {054301} (\bibinfo {year} {2019})}\BibitemShut {NoStop}%
\bibitem [{\citenamefont {Hamazaki}\ \emph
  {et~al.}(2019{\natexlab{a}})\citenamefont {Hamazaki}, \citenamefont
  {Kawabata},\ and\ \citenamefont {Ueda}}]{hamazaki2019non}%
  \BibitemOpen
  \bibfield  {author} {\bibinfo {author} {\bibfnamefont {R.}~\bibnamefont
  {Hamazaki}}, \bibinfo {author} {\bibfnamefont {K.}~\bibnamefont {Kawabata}},
  \ and\ \bibinfo {author} {\bibfnamefont {M.}~\bibnamefont {Ueda}},\ }\href
  {\doibase 10.1103/PhysRevLett.123.090603} {\bibfield  {journal} {\bibinfo
  {journal} {Phys. Rev. Lett.}\ }\textbf {\bibinfo {volume} {123}},\ \bibinfo
  {pages} {090603} (\bibinfo {year} {2019}{\natexlab{a}})}\BibitemShut
  {NoStop}%
\bibitem [{\citenamefont {Zhang}\ \emph {et~al.}(2020)\citenamefont {Zhang},
  \citenamefont {Tang}, \citenamefont {Lang}, \citenamefont {Yan},\ and\
  \citenamefont {Zhu}}]{zhang2020non}%
  \BibitemOpen
  \bibfield  {author} {\bibinfo {author} {\bibfnamefont {D.-W.}\ \bibnamefont
  {Zhang}}, \bibinfo {author} {\bibfnamefont {L.-Z.}\ \bibnamefont {Tang}},
  \bibinfo {author} {\bibfnamefont {L.-J.}\ \bibnamefont {Lang}}, \bibinfo
  {author} {\bibfnamefont {H.}~\bibnamefont {Yan}}, \ and\ \bibinfo {author}
  {\bibfnamefont {S.-L.}\ \bibnamefont {Zhu}},\ }\href@noop {} {\bibfield
  {journal} {\bibinfo  {journal} {Science China Physics, Mechanics \&
  Astronomy}\ }\textbf {\bibinfo {volume} {63}},\ \bibinfo {pages} {1}
  (\bibinfo {year} {2020})}\BibitemShut {NoStop}%
\bibitem [{\citenamefont {Markum}\ \emph {et~al.}(1999)\citenamefont {Markum},
  \citenamefont {Pullirsch},\ and\ \citenamefont {Wettig}}]{markum1999non}%
  \BibitemOpen
  \bibfield  {author} {\bibinfo {author} {\bibfnamefont {H.}~\bibnamefont
  {Markum}}, \bibinfo {author} {\bibfnamefont {R.}~\bibnamefont {Pullirsch}}, \
  and\ \bibinfo {author} {\bibfnamefont {T.}~\bibnamefont {Wettig}},\ }\href
  {\doibase 10.1103/PhysRevLett.83.484} {\bibfield  {journal} {\bibinfo
  {journal} {Phys. Rev. Lett.}\ }\textbf {\bibinfo {volume} {83}},\ \bibinfo
  {pages} {484} (\bibinfo {year} {1999})}\BibitemShut {NoStop}%
\bibitem [{\citenamefont {Verbaarschot}\ and\ \citenamefont
  {Wettig}(2000)}]{verbaarschot2000random}%
  \BibitemOpen
  \bibfield  {author} {\bibinfo {author} {\bibfnamefont {J.}~\bibnamefont
  {Verbaarschot}}\ and\ \bibinfo {author} {\bibfnamefont {T.}~\bibnamefont
  {Wettig}},\ }\href@noop {} {\bibfield  {journal} {\bibinfo  {journal} {Annual
  Review of Nuclear and Particle Science}\ }\textbf {\bibinfo {volume} {50}},\
  \bibinfo {pages} {343} (\bibinfo {year} {2000})}\BibitemShut {NoStop}%
\bibitem [{\citenamefont {Halasz}\ \emph
  {et~al.}(1997{\natexlab{a}})\citenamefont {Halasz}, \citenamefont {Jackson},\
  and\ \citenamefont {Verbaarschot}}]{halasz1997fermion}%
  \BibitemOpen
  \bibfield  {author} {\bibinfo {author} {\bibfnamefont {M.~A.}\ \bibnamefont
  {Halasz}}, \bibinfo {author} {\bibfnamefont {A.~D.}\ \bibnamefont {Jackson}},
  \ and\ \bibinfo {author} {\bibfnamefont {J.~J.~M.}\ \bibnamefont
  {Verbaarschot}},\ }\href {\doibase 10.1103/PhysRevD.56.5140} {\bibfield
  {journal} {\bibinfo  {journal} {Phys. Rev. D}\ }\textbf {\bibinfo {volume}
  {56}},\ \bibinfo {pages} {5140} (\bibinfo {year}
  {1997}{\natexlab{a}})}\BibitemShut {NoStop}%
\bibitem [{\citenamefont {Halasz}\ \emph
  {et~al.}(1997{\natexlab{b}})\citenamefont {Halasz}, \citenamefont {Osborn},\
  and\ \citenamefont {Verbaarschot}}]{halasz1997random}%
  \BibitemOpen
  \bibfield  {author} {\bibinfo {author} {\bibfnamefont {M.~A.}\ \bibnamefont
  {Halasz}}, \bibinfo {author} {\bibfnamefont {J.~C.}\ \bibnamefont {Osborn}},
  \ and\ \bibinfo {author} {\bibfnamefont {J.~J.~M.}\ \bibnamefont
  {Verbaarschot}},\ }\href {\doibase 10.1103/PhysRevD.56.7059} {\bibfield
  {journal} {\bibinfo  {journal} {Phys. Rev. D}\ }\textbf {\bibinfo {volume}
  {56}},\ \bibinfo {pages} {7059} (\bibinfo {year}
  {1997}{\natexlab{b}})}\BibitemShut {NoStop}%
\bibitem [{\citenamefont {Akemann}\ and\ \citenamefont
  {Bittner}(2006)}]{akemann2006unquenched}%
  \BibitemOpen
  \bibfield  {author} {\bibinfo {author} {\bibfnamefont {G.}~\bibnamefont
  {Akemann}}\ and\ \bibinfo {author} {\bibfnamefont {E.}~\bibnamefont
  {Bittner}},\ }\href {\doibase 10.1103/PhysRevLett.96.222002} {\bibfield
  {journal} {\bibinfo  {journal} {Phys. Rev. Lett.}\ }\textbf {\bibinfo
  {volume} {96}},\ \bibinfo {pages} {222002} (\bibinfo {year}
  {2006})}\BibitemShut {NoStop}%
\bibitem [{\citenamefont {Akemann}(2007)}]{akemann2007matrix}%
  \BibitemOpen
  \bibfield  {author} {\bibinfo {author} {\bibfnamefont {G.}~\bibnamefont
  {Akemann}},\ }\href@noop {} {\bibfield  {journal} {\bibinfo  {journal}
  {International Journal of Modern Physics A}\ }\textbf {\bibinfo {volume}
  {22}},\ \bibinfo {pages} {1077} (\bibinfo {year} {2007})}\BibitemShut
  {NoStop}%
\bibitem [{\citenamefont {Nishigaki}(2012{\natexlab{a}})}]{nishigaki2012level}%
  \BibitemOpen
  \bibfield  {author} {\bibinfo {author} {\bibfnamefont {S.~M.}\ \bibnamefont
  {Nishigaki}},\ }\href@noop {} {\bibfield  {journal} {\bibinfo  {journal}
  {Progress of theoretical physics}\ }\textbf {\bibinfo {volume} {128}},\
  \bibinfo {pages} {1283} (\bibinfo {year} {2012}{\natexlab{a}})}\BibitemShut
  {NoStop}%
\bibitem [{\citenamefont
  {Nishigaki}(2012{\natexlab{b}})}]{nishigaki2012universality}%
  \BibitemOpen
  \bibfield  {author} {\bibinfo {author} {\bibfnamefont {S.~M.}\ \bibnamefont
  {Nishigaki}},\ }\href {\doibase 10.1103/PhysRevD.86.114505} {\bibfield
  {journal} {\bibinfo  {journal} {Phys. Rev. D}\ }\textbf {\bibinfo {volume}
  {86}},\ \bibinfo {pages} {114505} (\bibinfo {year}
  {2012}{\natexlab{b}})}\BibitemShut {NoStop}%
\bibitem [{\citenamefont {Feinberg}(2011)}]{feinberg2011effective}%
  \BibitemOpen
  \bibfield  {author} {\bibinfo {author} {\bibfnamefont {J.}~\bibnamefont
  {Feinberg}},\ }\href {\doibase 10.1007/s10773-010-0604-y} {\bibfield
  {journal} {\bibinfo  {journal} {International Journal of Theoretical
  Physics}\ }\textbf {\bibinfo {volume} {50}},\ \bibinfo {pages} {1116}
  (\bibinfo {year} {2011})}\BibitemShut {NoStop}%
\bibitem [{\citenamefont {Kalish}\ \emph {et~al.}(2012)\citenamefont {Kalish},
  \citenamefont {Lin},\ and\ \citenamefont {Kottos}}]{kalish2012light}%
  \BibitemOpen
  \bibfield  {author} {\bibinfo {author} {\bibfnamefont {S.}~\bibnamefont
  {Kalish}}, \bibinfo {author} {\bibfnamefont {Z.}~\bibnamefont {Lin}}, \ and\
  \bibinfo {author} {\bibfnamefont {T.}~\bibnamefont {Kottos}},\ }\href
  {\doibase 10.1103/PhysRevA.85.055802} {\bibfield  {journal} {\bibinfo
  {journal} {Phys. Rev. A}\ }\textbf {\bibinfo {volume} {85}},\ \bibinfo
  {pages} {055802} (\bibinfo {year} {2012})}\BibitemShut {NoStop}%
\bibitem [{\citenamefont {Hatano}\ and\ \citenamefont
  {Feinberg}(2016)}]{hatano2016chebyshev}%
  \BibitemOpen
  \bibfield  {author} {\bibinfo {author} {\bibfnamefont {N.}~\bibnamefont
  {Hatano}}\ and\ \bibinfo {author} {\bibfnamefont {J.}~\bibnamefont
  {Feinberg}},\ }\href {\doibase 10.1103/PhysRevE.94.063305} {\bibfield
  {journal} {\bibinfo  {journal} {Phys. Rev. E}\ }\textbf {\bibinfo {volume}
  {94}},\ \bibinfo {pages} {063305} (\bibinfo {year} {2016})}\BibitemShut
  {NoStop}%
\bibitem [{\citenamefont {Mochizuki}\ and\ \citenamefont
  {Obuse}(2017)}]{mochizuki2017effects}%
  \BibitemOpen
  \bibfield  {author} {\bibinfo {author} {\bibfnamefont {K.}~\bibnamefont
  {Mochizuki}}\ and\ \bibinfo {author} {\bibfnamefont {H.}~\bibnamefont
  {Obuse}},\ }\href@noop {} {\bibfield  {journal} {\bibinfo  {journal}
  {Interdisciplinary Information Sciences}\ }\textbf {\bibinfo {volume} {23}},\
  \bibinfo {pages} {95} (\bibinfo {year} {2017})}\BibitemShut {NoStop}%
\bibitem [{\citenamefont {Bohigas}\ \emph {et~al.}(1984)\citenamefont
  {Bohigas}, \citenamefont {Giannoni},\ and\ \citenamefont
  {Schmit}}]{bohigas1984characterization}%
  \BibitemOpen
  \bibfield  {author} {\bibinfo {author} {\bibfnamefont {O.}~\bibnamefont
  {Bohigas}}, \bibinfo {author} {\bibfnamefont {M.~J.}\ \bibnamefont
  {Giannoni}}, \ and\ \bibinfo {author} {\bibfnamefont {C.}~\bibnamefont
  {Schmit}},\ }\href {\doibase 10.1103/PhysRevLett.52.1} {\bibfield  {journal}
  {\bibinfo  {journal} {Phys. Rev. Lett.}\ }\textbf {\bibinfo {volume} {52}},\
  \bibinfo {pages} {1} (\bibinfo {year} {1984})}\BibitemShut {NoStop}%
\bibitem [{\citenamefont {Ginibre}(1965)}]{ginibre1965statistical}%
  \BibitemOpen
  \bibfield  {author} {\bibinfo {author} {\bibfnamefont {J.}~\bibnamefont
  {Ginibre}},\ }\href@noop {} {\bibfield  {journal} {\bibinfo  {journal}
  {Journal of Mathematical Physics}\ }\textbf {\bibinfo {volume} {6}},\
  \bibinfo {pages} {440} (\bibinfo {year} {1965})}\BibitemShut {NoStop}%
\bibitem [{\citenamefont {Fyodorov}\ \emph {et~al.}(1997)\citenamefont
  {Fyodorov}, \citenamefont {Khoruzhenko},\ and\ \citenamefont
  {Sommers}}]{fyodorov1997almost}%
  \BibitemOpen
  \bibfield  {author} {\bibinfo {author} {\bibfnamefont {Y.~V.}\ \bibnamefont
  {Fyodorov}}, \bibinfo {author} {\bibfnamefont {B.~A.}\ \bibnamefont
  {Khoruzhenko}}, \ and\ \bibinfo {author} {\bibfnamefont {H.-J.}\ \bibnamefont
  {Sommers}},\ }\href {\doibase 10.1103/PhysRevLett.79.557} {\bibfield
  {journal} {\bibinfo  {journal} {Phys. Rev. Lett.}\ }\textbf {\bibinfo
  {volume} {79}},\ \bibinfo {pages} {557} (\bibinfo {year} {1997})}\BibitemShut
  {NoStop}%
\bibitem [{\citenamefont {Chalker}\ and\ \citenamefont
  {Mehlig}(1998)}]{chalker1998eigenvector}%
  \BibitemOpen
  \bibfield  {author} {\bibinfo {author} {\bibfnamefont {J.~T.}\ \bibnamefont
  {Chalker}}\ and\ \bibinfo {author} {\bibfnamefont {B.}~\bibnamefont
  {Mehlig}},\ }\href {\doibase 10.1103/PhysRevLett.81.3367} {\bibfield
  {journal} {\bibinfo  {journal} {Phys. Rev. Lett.}\ }\textbf {\bibinfo
  {volume} {81}},\ \bibinfo {pages} {3367} (\bibinfo {year}
  {1998})}\BibitemShut {NoStop}%
\bibitem [{\citenamefont {Bernard}\ and\ \citenamefont
  {LeClair}(2002)}]{bernard2002classification}%
  \BibitemOpen
  \bibfield  {author} {\bibinfo {author} {\bibfnamefont {D.}~\bibnamefont
  {Bernard}}\ and\ \bibinfo {author} {\bibfnamefont {A.}~\bibnamefont
  {LeClair}},\ }in\ \href@noop {} {\emph {\bibinfo {booktitle} {Statistical
  Field Theories}}}\ (\bibinfo  {publisher} {Springer},\ \bibinfo {year}
  {2002})\ pp.\ \bibinfo {pages} {207--214}\BibitemShut {NoStop}%
\bibitem [{\citenamefont {Shukla}(2001)}]{shukla2001non}%
  \BibitemOpen
  \bibfield  {author} {\bibinfo {author} {\bibfnamefont {P.}~\bibnamefont
  {Shukla}},\ }\href {\doibase 10.1103/PhysRevLett.87.194102} {\bibfield
  {journal} {\bibinfo  {journal} {Phys. Rev. Lett.}\ }\textbf {\bibinfo
  {volume} {87}},\ \bibinfo {pages} {194102} (\bibinfo {year}
  {2001})}\BibitemShut {NoStop}%
\bibitem [{\citenamefont {Garc\'{\i}a-Garc\'{\i}a}\ \emph
  {et~al.}(2002)\citenamefont {Garc\'{\i}a-Garc\'{\i}a}, \citenamefont
  {Nishigaki},\ and\ \citenamefont {Verbaarschot}}]{garcia2002critical}%
  \BibitemOpen
  \bibfield  {author} {\bibinfo {author} {\bibfnamefont {A.~M.}\ \bibnamefont
  {Garc\'{\i}a-Garc\'{\i}a}}, \bibinfo {author} {\bibfnamefont {S.~M.}\
  \bibnamefont {Nishigaki}}, \ and\ \bibinfo {author} {\bibfnamefont
  {J.~J.~M.}\ \bibnamefont {Verbaarschot}},\ }\href {\doibase
  10.1103/PhysRevE.66.016132} {\bibfield  {journal} {\bibinfo  {journal} {Phys.
  Rev. E}\ }\textbf {\bibinfo {volume} {66}},\ \bibinfo {pages} {016132}
  (\bibinfo {year} {2002})}\BibitemShut {NoStop}%
\bibitem [{\citenamefont {Ahmed}(2003)}]{ahmed2003ensemble}%
  \BibitemOpen
  \bibfield  {author} {\bibinfo {author} {\bibfnamefont {Z.}~\bibnamefont
  {Ahmed}},\ }\href@noop {} {\bibfield  {journal} {\bibinfo  {journal} {Physics
  Letters A}\ }\textbf {\bibinfo {volume} {308}},\ \bibinfo {pages} {140}
  (\bibinfo {year} {2003})}\BibitemShut {NoStop}%
\bibitem [{\citenamefont {Feinberg}(2006)}]{feinberg2006non}%
  \BibitemOpen
  \bibfield  {author} {\bibinfo {author} {\bibfnamefont {J.}~\bibnamefont
  {Feinberg}},\ }\href@noop {} {\bibfield  {journal} {\bibinfo  {journal}
  {Journal of Physics A: Mathematical and General}\ }\textbf {\bibinfo {volume}
  {39}},\ \bibinfo {pages} {10029} (\bibinfo {year} {2006})}\BibitemShut
  {NoStop}%
\bibitem [{\citenamefont {Akemann}\ \emph
  {et~al.}(2009{\natexlab{a}})\citenamefont {Akemann}, \citenamefont
  {Phillips},\ and\ \citenamefont {Shifrin}}]{akemann2009gap}%
  \BibitemOpen
  \bibfield  {author} {\bibinfo {author} {\bibfnamefont {G.}~\bibnamefont
  {Akemann}}, \bibinfo {author} {\bibfnamefont {M.}~\bibnamefont {Phillips}}, \
  and\ \bibinfo {author} {\bibfnamefont {L.}~\bibnamefont {Shifrin}},\
  }\href@noop {} {\bibfield  {journal} {\bibinfo  {journal} {Journal of
  Mathematical Physics}\ }\textbf {\bibinfo {volume} {50}},\ \bibinfo {pages}
  {063504} (\bibinfo {year} {2009}{\natexlab{a}})}\BibitemShut {NoStop}%
\bibitem [{\citenamefont {Akemann}\ \emph
  {et~al.}(2009{\natexlab{b}})\citenamefont {Akemann}, \citenamefont {Bittner},
  \citenamefont {Phillips},\ and\ \citenamefont {Shifrin}}]{akemann2009wigner}%
  \BibitemOpen
  \bibfield  {author} {\bibinfo {author} {\bibfnamefont {G.}~\bibnamefont
  {Akemann}}, \bibinfo {author} {\bibfnamefont {E.}~\bibnamefont {Bittner}},
  \bibinfo {author} {\bibfnamefont {M.~J.}\ \bibnamefont {Phillips}}, \ and\
  \bibinfo {author} {\bibfnamefont {L.}~\bibnamefont {Shifrin}},\ }\href
  {\doibase 10.1103/PhysRevE.80.065201} {\bibfield  {journal} {\bibinfo
  {journal} {Phys. Rev. E}\ }\textbf {\bibinfo {volume} {80}},\ \bibinfo
  {pages} {065201(R)} (\bibinfo {year} {2009}{\natexlab{b}})}\BibitemShut
  {NoStop}%
\bibitem [{\citenamefont {Joglekar}\ and\ \citenamefont
  {Karr}(2011)}]{joglekar2011level}%
  \BibitemOpen
  \bibfield  {author} {\bibinfo {author} {\bibfnamefont {Y.~N.}\ \bibnamefont
  {Joglekar}}\ and\ \bibinfo {author} {\bibfnamefont {W.~A.}\ \bibnamefont
  {Karr}},\ }\href {\doibase 10.1103/PhysRevE.83.031122} {\bibfield  {journal}
  {\bibinfo  {journal} {Phys. Rev. E}\ }\textbf {\bibinfo {volume} {83}},\
  \bibinfo {pages} {031122} (\bibinfo {year} {2011})}\BibitemShut {NoStop}%
\bibitem [{\citenamefont {Bohigas}\ and\ \citenamefont
  {Pato}(2013)}]{bohigas2013non}%
  \BibitemOpen
  \bibfield  {author} {\bibinfo {author} {\bibfnamefont {O.}~\bibnamefont
  {Bohigas}}\ and\ \bibinfo {author} {\bibfnamefont {M.~P.}\ \bibnamefont
  {Pato}},\ }\href@noop {} {\bibfield  {journal} {\bibinfo  {journal} {AIP
  Advances}\ }\textbf {\bibinfo {volume} {3}},\ \bibinfo {pages} {032130}
  (\bibinfo {year} {2013})}\BibitemShut {NoStop}%
\bibitem [{\citenamefont {Graefe}\ \emph {et~al.}(2015)\citenamefont {Graefe},
  \citenamefont {Mudute-Ndumbe},\ and\ \citenamefont
  {Taylor}}]{graefe2015random}%
  \BibitemOpen
  \bibfield  {author} {\bibinfo {author} {\bibfnamefont {E.-M.}\ \bibnamefont
  {Graefe}}, \bibinfo {author} {\bibfnamefont {S.}~\bibnamefont
  {Mudute-Ndumbe}}, \ and\ \bibinfo {author} {\bibfnamefont {M.}~\bibnamefont
  {Taylor}},\ }\href@noop {} {\bibfield  {journal} {\bibinfo  {journal}
  {Journal of Physics A: Mathematical and Theoretical}\ }\textbf {\bibinfo
  {volume} {48}},\ \bibinfo {pages} {38FT02} (\bibinfo {year}
  {2015})}\BibitemShut {NoStop}%
\bibitem [{\citenamefont {Hamazaki}\ \emph
  {et~al.}(2019{\natexlab{b}})\citenamefont {Hamazaki}, \citenamefont
  {Kawabata}, \citenamefont {Kura},\ and\ \citenamefont
  {Ueda}}]{hamazaki2019threefold}%
  \BibitemOpen
  \bibfield  {author} {\bibinfo {author} {\bibfnamefont {R.}~\bibnamefont
  {Hamazaki}}, \bibinfo {author} {\bibfnamefont {K.}~\bibnamefont {Kawabata}},
  \bibinfo {author} {\bibfnamefont {N.}~\bibnamefont {Kura}}, \ and\ \bibinfo
  {author} {\bibfnamefont {M.}~\bibnamefont {Ueda}},\ }\href@noop {} {\bibfield
   {journal} {\bibinfo  {journal} {arXiv preprint arXiv:1904.13082}\ }
  (\bibinfo {year} {2019}{\natexlab{b}})}\BibitemShut {NoStop}%
\bibitem [{\citenamefont {Tzortzakakis}\ \emph {et~al.}(2020)\citenamefont
  {Tzortzakakis}, \citenamefont {Makris},\ and\ \citenamefont
  {Economou}}]{tzortzakakis2020non}%
  \BibitemOpen
  \bibfield  {author} {\bibinfo {author} {\bibfnamefont {A.~F.}\ \bibnamefont
  {Tzortzakakis}}, \bibinfo {author} {\bibfnamefont {K.~G.}\ \bibnamefont
  {Makris}}, \ and\ \bibinfo {author} {\bibfnamefont {E.~N.}\ \bibnamefont
  {Economou}},\ }\href {\doibase 10.1103/PhysRevB.101.014202} {\bibfield
  {journal} {\bibinfo  {journal} {Phys. Rev. B}\ }\textbf {\bibinfo {volume}
  {101}},\ \bibinfo {pages} {014202} (\bibinfo {year} {2020})}\BibitemShut
  {NoStop}%
\bibitem [{\citenamefont {Mehta}(2004)}]{mehta2004random}%
  \BibitemOpen
  \bibfield  {author} {\bibinfo {author} {\bibfnamefont {M.~L.}\ \bibnamefont
  {Mehta}},\ }\href@noop {} {\emph {\bibinfo {title} {Random matrices}}}\
  (\bibinfo  {publisher} {Elsevier},\ \bibinfo {year} {2004})\BibitemShut
  {NoStop}%
\bibitem [{\citenamefont {Dyson}(1953)}]{dyson1953the}%
  \BibitemOpen
  \bibfield  {author} {\bibinfo {author} {\bibfnamefont {F.~J.}\ \bibnamefont
  {Dyson}},\ }\href {\doibase 10.1103/PhysRev.92.1331} {\bibfield  {journal}
  {\bibinfo  {journal} {Phys. Rev.}\ }\textbf {\bibinfo {volume} {92}},\
  \bibinfo {pages} {1331} (\bibinfo {year} {1953})}\BibitemShut {NoStop}%
\bibitem [{\citenamefont {Dunne}\ and\ \citenamefont
  {Feinberg}(1998)}]{dunne1998self}%
  \BibitemOpen
  \bibfield  {author} {\bibinfo {author} {\bibfnamefont {G.~V.}\ \bibnamefont
  {Dunne}}\ and\ \bibinfo {author} {\bibfnamefont {J.}~\bibnamefont
  {Feinberg}},\ }\href {\doibase 10.1103/PhysRevD.57.1271} {\bibfield
  {journal} {\bibinfo  {journal} {Phys. Rev. D}\ }\textbf {\bibinfo {volume}
  {57}},\ \bibinfo {pages} {1271} (\bibinfo {year} {1998})}\BibitemShut
  {NoStop}%
\bibitem [{\citenamefont {Theodorou}\ and\ \citenamefont
  {Cohen}(1976)}]{theodorou1976extended}%
  \BibitemOpen
  \bibfield  {author} {\bibinfo {author} {\bibfnamefont {G.}~\bibnamefont
  {Theodorou}}\ and\ \bibinfo {author} {\bibfnamefont {M.~H.}\ \bibnamefont
  {Cohen}},\ }\href {\doibase 10.1103/PhysRevB.13.4597} {\bibfield  {journal}
  {\bibinfo  {journal} {Phys. Rev. B}\ }\textbf {\bibinfo {volume} {13}},\
  \bibinfo {pages} {4597} (\bibinfo {year} {1976})}\BibitemShut {NoStop}%
\bibitem [{\citenamefont {Eggarter}\ and\ \citenamefont
  {Riedinger}(1978)}]{eggarter1978singular}%
  \BibitemOpen
  \bibfield  {author} {\bibinfo {author} {\bibfnamefont {T.~P.}\ \bibnamefont
  {Eggarter}}\ and\ \bibinfo {author} {\bibfnamefont {R.}~\bibnamefont
  {Riedinger}},\ }\href {\doibase 10.1103/PhysRevB.18.569} {\bibfield
  {journal} {\bibinfo  {journal} {Phys. Rev. B}\ }\textbf {\bibinfo {volume}
  {18}},\ \bibinfo {pages} {569} (\bibinfo {year} {1978})}\BibitemShut
  {NoStop}%
\bibitem [{\citenamefont {Titov}\ \emph {et~al.}(2001)\citenamefont {Titov},
  \citenamefont {Brouwer}, \citenamefont {Furusaki},\ and\ \citenamefont
  {Mudry}}]{titov2001fokker}%
  \BibitemOpen
  \bibfield  {author} {\bibinfo {author} {\bibfnamefont {M.}~\bibnamefont
  {Titov}}, \bibinfo {author} {\bibfnamefont {P.~W.}\ \bibnamefont {Brouwer}},
  \bibinfo {author} {\bibfnamefont {A.}~\bibnamefont {Furusaki}}, \ and\
  \bibinfo {author} {\bibfnamefont {C.}~\bibnamefont {Mudry}},\ }\href
  {\doibase 10.1103/PhysRevB.63.235318} {\bibfield  {journal} {\bibinfo
  {journal} {Phys. Rev. B}\ }\textbf {\bibinfo {volume} {63}},\ \bibinfo
  {pages} {235318} (\bibinfo {year} {2001})}\BibitemShut {NoStop}%
\end{thebibliography}%

\end{document}